\documentclass[aps,showpacs,pra, onecolumn,  superscriptaddress,notitlepage]{revtex4-1}

\usepackage[utf8]{inputenc}
\usepackage{exscale}
\usepackage{bbm}
\usepackage{graphicx}
\usepackage{amsmath}
\usepackage{latexsym}
\usepackage{amsfonts}
\usepackage{amssymb}
\usepackage{times}
\usepackage[T1]{fontenc}
\usepackage{amsthm}
\usepackage{enumitem}
\usepackage{bbold}
\usepackage{color}
\usepackage[colorlinks=true,citecolor=blue,urlcolor=blue]{hyperref}
\usepackage{mathtools}
\usepackage{changes}
\usepackage[normalem]{ulem}
\usepackage{subcaption}
\usepackage{braket}
\usepackage{xspace}
\usepackage{textcomp}

\newcommand{\eqnref}[1]{Eq.~(\ref{#1})}
\newcommand{\figref}[1]{Fig.~\ref{#1}}

\newcommand{\secref}[1]{Section~\ref{#1}}
\newcommand{\appref}[1]{Appendix~\ref{#1}}

\DeclareMathOperator*{\argmax}{arg\,max}

\newcommand{\eg}{\unskip\space\textit{e.g.}\xspace}
\newcommand{\ie}{\unskip\space\textit{i.e.}\xspace}

\definecolor{jordi}{rgb}{0.1,0.1,0.5}

\definecolor{borja}{rgb}{0.9,0.2,0}

\begin{document}

\author{Borja Requena}
\affiliation{ICFO - Institut de Ciencies Fotoniques, The Barcelona Institute of Science and Technology, Av. Carl Friedrich Gauss 3, 08860 Castelldefels (Barcelona), Spain}

\author{Gorka Muñoz-Gil}
\affiliation{ICFO - Institut de Ciencies Fotoniques, The Barcelona Institute of Science and Technology, Av. Carl Friedrich Gauss 3, 08860 Castelldefels (Barcelona), Spain}

\author{Maciej Lewenstein}
\affiliation{ICFO - Institut de Ciencies Fotoniques, The Barcelona Institute of Science and Technology, Av. Carl Friedrich Gauss 3, 08860 Castelldefels (Barcelona), Spain}
\affiliation{ICREA, Pg. Llu\'is Companys 23, 08010 Barcelona, Spain}

\author{Vedran Dunjko}
\affiliation{LIACS, Leiden University, Niels Bohrweg 1, 2333 CA Leiden, Netherlands}

\author{Jordi Tura}
\affiliation{Max-Planck-Institut f\"ur Quantenoptik, Hans-Kopfermann-Str. 1, 85748 Garching, Germany}
\affiliation{Instituut-Lorentz, Universiteit Leiden, P.O. Box 9506, 2300 RA Leiden, The Netherlands}

\title{Certificates of quantum many-body properties assisted by machine learning}

\begin{abstract}
Computationally intractable tasks are often encountered in physics and optimization. Such tasks often comprise a cost function to be optimized over a so-called feasible set, which is specified by a set of constraints. This may yield, in general, to difficult and non-convex optimization tasks. A number of standard methods are used to tackle such problems: variational approaches focus on parameterizing a subclass of solutions within the feasible set; in contrast, relaxation techniques have been proposed to approximate it from outside, thus complementing the variational approach by providing ultimate bounds to the global optimal solution.
In this work, we propose a novel approach combining the power of relaxation techniques with deep reinforcement learning in order to find the best possible bounds within a limited computational budget. We illustrate the viability of the method in the context of finding the ground state energy of many-body quantum systems, a paradigmatic problem in quantum physics. We benchmark our approach against other classical optimization algorithms such as breadth-first search or Monte-Carlo, and we characterize the effect of transfer learning. We find the latter may be indicative of phase transitions, with a completely autonomous approach. Finally, we provide tools to generalize the approach to other common applications in the field of quantum information processing.

\end{abstract}

\flushbottom
\maketitle

\section{Introduction}
\begin{figure}
    \centering
    \begin{subfigure}{.5\textwidth}
      \centering
      \includegraphics[width=.62\linewidth]{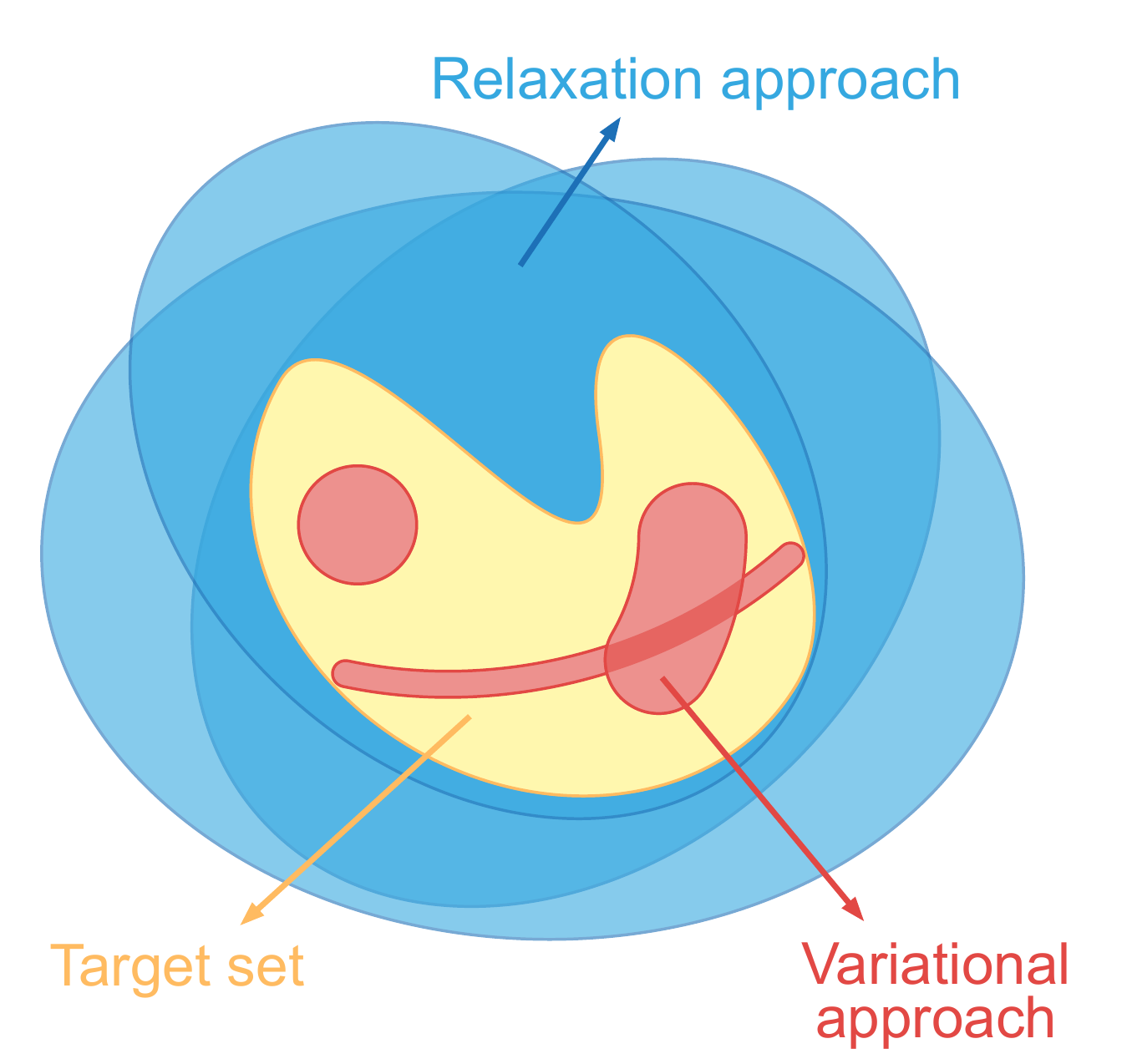}
      \caption{Schematic representation of the optimization task. The task is to optimize a function over a hard to characterize set (yellow set). The variational approach allows to parameterize subsets within the set of interest (different red sets). Different parameterizations yield different subsets that are more or less convenient depending on the task. Relaxation techniques allow to efficiently represent larger sets than the one of interest (different blue sets) exploiting, for instance, convexity or linearity. Neither different variational approaches nor different relaxations need to be contained into one another, so the sets they represent are incomparable in general.}
      \label{fig:sets1}
    \end{subfigure}%
    \begin{subfigure}{.4\textwidth}
      \centering
      \includegraphics[width=.6\linewidth]{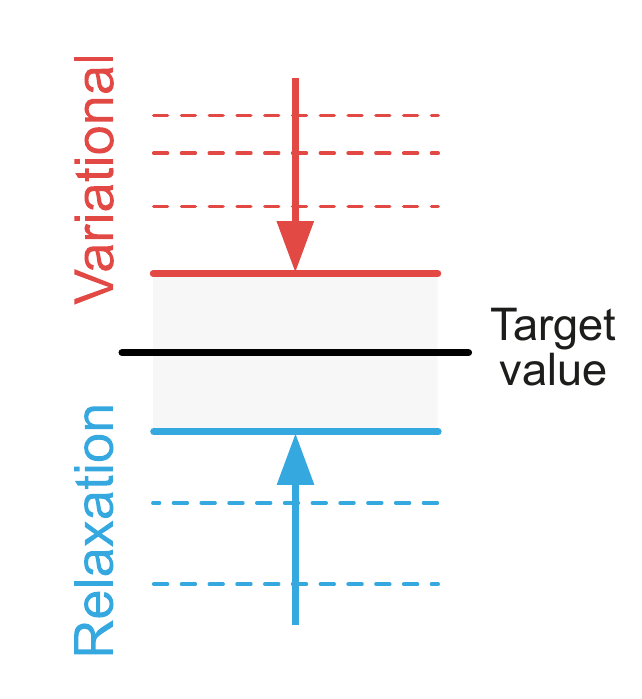}
      \caption{Values of the objective function. In black, the optimal unknown value. In red, the different minima obtained by variational methods. The smaller the value, the better the bound. In blue, different minima obtained by relaxation techniques. The greater the value, the more accurate their associated certificate. In grey, the uncertainty region where the optimal solution lies, given by the best variational and the best certificate obtained so far.}
      \label{fig:sets2}
    \end{subfigure}%
    \caption{Interpretation of exact solutions, variational solutions and certificates obtained by relaxation techniques.}
    \label{fig:sets}
\end{figure}

Computationally intractable tasks naturally appear at the core of physics and optimization. There exist two paradigmatic approaches to address them (see~\figref{fig:sets}). The first one is based on the variational ansatz: here one parameterizes a family of solutions with the hope that it contains, at least, a good approximation to the optimal one. The more complex the ansatz is, the higher are the chances to approximately represent the optimal solution, but, at the same time, the more computationally expensive the search becomes. Since the parameterized families of solutions are such that they satisfy the problem's constraints, variational approaches are sub-optimal by construction (they might not even contain the global optimum), thus providing a bound from one side. The second approach is based on relaxation methods. In this case, rather than focusing on finding an example, these methods look for a mathematical proof: by optimizing over a superset of the feasible set, one can write an easier optimization task. For instance, the relaxed set may be obtained by lifting some of the constraints or restrictions that define the feasible set, and that may simplify the optimization. This approach thus yields a bound from the other side. Such a proof is often referred to as a certificate. In order to obtain simpler certificates, the space of solutions is normally extended, \eg to include non-physical states, with the goal to imbue the feasible set with a desirable property, such as convexity. This process constitutes a so-called relaxation. A good relaxation makes the proof easier to obtain, for instance, by using optimization methods such as semidefinite programming (SdP). The combination of the two approaches yields an upper and lower bound; \ie, an uncertainty interval around the optimal solution, as illustrated in~\figref{fig:sets2}. In addition, both approaches are typically fine-tunable in terms of the required computational resources: physical insight has motivated many variational approaches that efficiently achieve good bounds in some cases. Similarly, one can also look for shorter, uncomplicated mathematical proofs in order to learn about the structure of the optimization task.

In a quantum context, the variational ansatz has found tremendous success in areas so diverse as quantum chemistry \cite{KandalaNature2017, KandalaNature2019, PeruzzoNatComms2014, LanyonNatChem2010, HempelPRX2018, OMalleyPRX2016, OBriennpjQI2019}, condensed matter \cite{WhitePRL1992, WhitePRB1993, VerstraetePRL2004, DaleyJSMTE2004, OrusAoP2014, Bravo-Prieto2020}, and quantum machine learning \cite{BiamonteNature2017, DunjkoROPP2018}. In the so-called noisy, intermediate-scale quantum (NISQ) era \cite{PreskillQuantum2018}, the variational ansatz is the main pillar upon which quantum algorithms such as quantum approximate optimization algorithms \cite{Farhi2014, KokailNature2019, Crooks2018, ZhouPRX2020, Arute2020} and variational quantum eigensolvers \cite{Herasymenko2019, Garcia-Saez2018, Bravo-Prieto2020, SagastizabalPRA2019, TuraCEUR2020, BenedettiNPJ2019} rest. However, in all these cases, variational solutions are suboptimal by construction and, even if they happen to actually represent the optimal one, additional methods are required to prove such a claim.
By increasing the size of the parameter space, one can of course represent better solutions, but this comes at the cost of demanding more computational resources. Furthermore, the distance between the best solution found and the optimal one is unknown in general. This information is of paramount practical importance in order to decide whether it is worth to spend more resources in looking for a better solution or stop the search.
 
On the other hand, relaxation techniques have been widely used in quantum information processing (QIP) since its dawn (see \figref{fig:sets}). Perhaps, the most paradigmatic example in the context of entanglement theory is the Peres criterion, which is a relaxation from the set of separable states to the set of states that are positive under partial transposition (PPT) \cite{PeresPRL1996}. The membership problem in the separable set was shown to be NP-hard \cite{Gurvits2003}, whereas checking whether the PPT criterion is violated is very simple, yielding one of the simplest ways to show a quantum state is entangled.
However, not all quantum states in the PPT set are separable; \ie, the relaxed set contains states that are both entangled and PPT \cite{HorodeckiPLA1996}. A systematic way to strengthen the PPT criterion is via symmetric extensions \cite{DohertyPRA2004, Marconi2020}, which are families of increasingly better, albeit increasingly costly, SdP-based certificates. In the device-independent version of QIP \cite{AcinDIQKD}, following a similar philosophy, relaxation techniques have also played a major role. For instance, in cryptographic security proofs, one needs to be safe against all possible quantum attacks, which are very difficult to characterize, therefore motivating research for supraquantum theories that are more tractable analytically \cite{GallegoNatComms2013, AugusiakPRA2014}. Indeed, in the quest for the characterization of the set of quantum correlations \cite{Slofstra2017}, several operationally simple, outer approximations have been proposed \cite{PR94, BrassardPRL2006, LindenPRL2007, NavascuesPRSA2010, PawlowskiNature2009, FritzNatComms2013, GallegoPRL2011, NavascuesNatComms2015}, as well as systematic relaxations via SdP relaxations \cite{NavascuesPRL2007, NavascuesNJP2008, PironioSIAM2010}. Many variations over this method have been developed in different scenarios \cite{YangPRAR2013, BudroniPRL2013, PozasKerstjensPRL2019, AloyPRL2019, TuraPRA2019, ChenPRL2016, ChenPRA2018, Chen2020}, \eg, their commutative counterpart \cite{LasserreSIAM2001, ParriloBook2013} has been studied in various contexts related to local hidden variable theories \cite{BaccariPRX2017, FadelPRL2017} and classical spin models \cite{BaccariPRR2020}.

Not surprisingly, simpler proofs may be easier to obtain, although they may also yield looser bounds. At the same time, some proofs may be more elegant/smarter than others of similar complexity, yielding a better bound while using similar computational resources. The latter ones generally exploit useful properties of the system, such as the existence of symmetries. This has been paramount in self-testing protocols based on operator-sums-of-squares (OSOS) decompositions, which are, again, obtained via a SdP. One of the main difficulties encountered in finding OSOS is to find analytical proofs which are simple enough to be manageable \cite{BampsPRA2015, SATWAP, KaniewskiQuantum2019, AugusiakNJP2019, BASTA}. In other words, it is paramount to find, among all possible relaxations of the original problem, the best trade-off between accuracy and simplicity. However, a successful search often relies on specific insight about the problem at hand. Conversely, the analysis of an efficient proof is more likely to reveal useful insight about the system's properties.

In last years, machine learning approaches have shown great success at solving combinatorial optimization problems like the one previously described~\cite{BengioEJOR2021}. Different architectures have been proposed, from supervised learning of neural networks~\cite{Vinyals2015} to unsupervised approaches over graphs~\cite{Karalias2020}. Closer to the problem proposed on this paper, the former have been used to ease the solution of SdP relaxations~\cite{BalteanTechRep2018}. Another common approach for such problems has been the use of reinforcement learning~\cite{SuttonBook2018, Mazyavkina2020}. While traditional algorithms rely on heuristics and specific insight about the nature of the problem, machine learning approaches are able to solve many of these without any prior knowledge and faster than those. In the same line, machine learning approaches of all kinds have been lately applied to several problems in physics~\cite{CarleoRMP2019}.

In this work, we propose a method to systematically search for an optimal relaxation within a given computational budget, using reinforcement learning (RL) techniques. We propose a scheme in which an agent has access to a black box that computes the relaxation of the problem by solving an SdP (see \figref{fig:RL_framework}). The agent can increase or decrease the relaxation level, observing an output that depends on both the computational cost and the quality of the obtained certificate. We illustrate the procedure in the context of finding the ground state energy of local Hamiltonians. Our results show that even for very simple scenarios we find counter-intuitive optimal relaxations. Then, we compare our RL approach to other optimization algorithms and, finally, we show how to use transfer learning in the proposed framework.

Applying RL to obtain useful certificates can be seen as a meta-algorithm with a wide range of applicability. In this work we shall present it through a running case study, without hindering its more general flavor. In \secref{sec:Generalization} we discuss how the same principles, as presented here, apply to diverse areas of quantum information processing.

The paper is structured as follows: In \secref{sec:prelim} we describe the optimization task that we consider as our running example. We discuss the natural ways to build a relaxation out of this task, by imposing a set of constraints. In \secref{sec:constraintspace} we introduce the constraint space for the agent and in \secref{sec:constraintopt} we introduce the optimization framework. In the latter, we define the state space, actions and rewards for the agent. We present our main results in \secref{sec:results}. In particular, we devote \secref{sec:benchmarking} to benchmarking the proposed approach and \secref{sec:transfer} to characterizing the possibilities of transfer learning. Then, we discuss some particular cases of interest in \secref{sec:Particularcases} and we discuss how our framework naturally applies to various relevant problems in quantum information in \secref{sec:Generalization}. Finally, we conclude in \secref{sec:concl}.

\section{Preliminaries}
\label{sec:prelim}
This section explains the methods to systematically build certificates, which we are going to consider throughout the paper. These certificates are based on the optimal solution of a semidefinite program. The optimality of the SdP solution or, at least, a valid bound for a certificate follows from strong or weak duality properties, respectively (cf. Appendix \ref{app:SdPBasics}). This methodology will be incorporated into the reinforcement learning procedure in Section \ref{sec:constraintopt} as a black box module.
In the interest of simplicity, throughout all the paper we shall consider a running example. While this does not restrict the applicability of our work to other areas in quantum information (see \secref{sec:Generalization}), it shall certainly ease the exposition.

Let us therefore fix an optimization task, which is to find the ground state energy $E_0$ of a quantum local Hamiltonian
\begin{equation}
    H = \sum_{i=1}^m H_i.
    \label{eq:localH}
\end{equation}
The Hamiltonian $H$ acts on $n$ qubits, and it is a sum of terms $H_i$, each of which acts on at most $k=O(1)$ qubits. The sum \eqnref{eq:localH} has therefore $m=O(\mathrm{poly}(n))$ terms. The support of $H_i$, denoted $\mathrm{supp}(H_i)$ is the set of qubits where $H_i$ acts non-trivially. The supports of the different $H_i$ may overlap; \ie, $\mathrm{supp}(H_i)\cap \mathrm{supp}(H_j)$ may not be empty.

To find $E_0$, a possibility is to directly construct a quantum state that has $E_0$ energy with respect to $H$. Therefore, a first possible approach is to parameterize a family of quantum states $\ket{\psi(\boldsymbol{\theta})}$ exploiting some known properties of $H$.  We can safely assume the parameterization yields a valid (\ie, normalized) quantum state for any value of the parameters $\boldsymbol{\theta}$. Additionally, by construction, $\bra{\psi(\boldsymbol{\theta})}H\ket{\psi({\boldsymbol{\theta}})} \geq E_0$ for all $\boldsymbol{\theta}$. Let us denote
\begin{equation}
    \gamma = \min_{\boldsymbol{\theta}} \bra{\psi(\boldsymbol{\theta})}H\ket{\psi({\boldsymbol{\theta}})},
\end{equation}
which satisfies $\gamma \geq E_0$ by construction. An example of such a parameterization would be to describe $\ket{\psi(\boldsymbol{\theta})}$ as a tensor network contraction, which exploits the locality properties of $H$, limiting the entanglement present in its ground state \cite{OrusAoP2014,VerstraeteAdP2008,SchuchAnnPhys2010,SchuchPRA2010,ZhouPRX2020a}.

Complexity theory results (in particular, QMA-hardness) strongly suggest that finding, or even approximating, the ground state energy of a local Hamiltonian is a hard task, even for a quantum computer \cite{KempeQIC2003, KempeSIAM2006, AharonovCMP2009}. Furthermore, this hardness persists in physically relevant instances \cite{SchuchNatPhys2009}. Notice that, even if we found the actual solution $\ket{\psi(\boldsymbol{\theta})}$, we cannot prove, solely from that, that it is the global minimum \cite{Czartowski2018}.

It is therefore highly desirable to obtain a bound from the other side; \ie, a value $\beta$ for which one can prove $E_0 \geq \beta$. This would guarantee $E_0 \in [\beta, \gamma]$ and, thus, help determine whether it is worth to refine the search depending on $|\gamma - \beta| < \varepsilon$. However, for a proof of the type $E_0 \geq \beta$, constructing an example $\ket{\psi(\boldsymbol{\theta})}$ is not good enough. We need a proof that is satisfied by all valid quantum states and, possibly, a larger set, as long as it makes the proof simpler. Such a proof is referred to as a certificate, and it is typically obtained by numerical means. SdP is a natural tool to obtain such certificates upon which we capitalize in our work.

A common technique to construct a relaxation for the local Hamiltonian problem is via the triangle inequality \cite{AndersonPR1951, TarrachPRB1990, ChandranPRL2007, AletPRLComment2008}:
\begin{equation}
    \min_{\rho} \mathrm{Tr}[\rho H] \geq \sum_{i} \min_{\rho_i} \mathrm{Tr}[\rho_i \hat{H}_i],
    \label{eq:Triangle}
\end{equation}
where $\rho$ and $\rho_i$ are density matrices acting on the support of $H$ and $H_i$ respectively. Note that $i$ refers to a Hamiltonian term and it has nothing to do with the $i$-th party. Furthermore, in \eqnref{eq:Triangle}, the $\hat{H}_i$ are sums of some local terms $H_j$ of \eqnref{eq:localH}, grouped so that $\mathrm{supp}(\hat{H}_i)$ is as large as possible while still allowing for computation of their minimal eigenvalue. This size obviously depends on the available computational resources.

Let us observe that the RHS in \eqnref{eq:Triangle} is a sum of minima, where each minimization is carried out independently. Due to this independence, in general, it is not the case that different $\rho_i$ are mutually compatible; \ie, that there exists a global state $\rho$ such that each $\rho_i$ is the corresponding partial trace of $\rho$. The converse is true, however: every valid quantum state $\rho$ has an associated set of partial traces $\rho_i$, but given a set of $\rho_i$, a global $\rho$ may not exist. This is what proves the inequality \eqnref{eq:Triangle}.

The minimization of the RHS of \eqnref{eq:Triangle} is equivalent to solving the following SdP (cf. \appref{app:SdPBasics}):
\begin{equation}
    \begin{array}{llr}
    \beta_{\emptyset}:=&\min_{\{\rho_i\}}& \sum_{i} \mathrm{Tr}[\rho_i \hat{H}_i]\\
    &\mathrm{s.t.}& \rho_{i} \succeq 0\\
    &&\mathrm{Tr}[\rho_i] = 1.
    \end{array}
    \label{eq:TrivialSdP}
\end{equation}

Since there is no mutual compatibility enforced among the $\rho_i$, and each is treated independently, the triangle inequality \eqnref{eq:Triangle} constitutes a trivial relaxation. A natural way to strengthen the relaxation is to impose further restrictions on the collection of possible $\rho_i$, in such a way that any quantum state would also satisfy them. The strongest restriction possible is to directly ask that $\{\rho_i\}$ come from a global quantum state. Unfortunately, this would be equivalent to finding the value of $E_0$, which is QMA-complete. Furthermore, it is strongly connected to solving the so-called quantum marginal problem (QMP), which is also QMA-complete \cite{KempeQIC2003, KempeSIAM2006, AharonovCMP2009}. The QMP has been solved completely in very rare instances, such as the global state being symmetric \cite{AloyNJP2021} or for the case of one-body marginals \cite{WalterScience2013}. Nevertheless, the SdP based formulation \eqnref{eq:TrivialSdP} motivates a hierarchy of relaxations based on solving the QMP up to some degree of compatibility.

\subsection{Constructing tighter certificates}

In order to build certificates that yield a tighter bound than that of the triangle inequality, our first observation is that the set $\{\rho_i\}$ does not need fulfill any mutual compatibility constraint. It would be natural to expect that, at least, the partial traces on different supports' intersection match. This will reduce the space of solutions, provided that $\{\rho_i\}$ must fulfill additional conditions. Therefore, since the minimization is over a smaller set, its result can only be a tighter bound.

Hence, the first level of compatibility we might want to ask for is that $\rho_i$ and $\rho_j$ yield the same reduced density matrix (RDM) on their common support, which we shall denote $\rho_{i \wedge j}$:
\begin{equation}
    \mathrm{Tr}_{\mathrm{supp}(\rho_j)^c}[\rho_i] = \mathrm{Tr}_{\mathrm{supp}(\rho_i)^c}[\rho_j] \equiv \rho_{i \wedge j}.
\end{equation}
Here, the partial trace $\mathrm{Tr}_S(\cdot)$ denotes that we eliminate subsystem $S$ and the superindex $c$ indicates the complementary set. Thus, $\mathrm{Tr}_{S^c}$ produces the RDM acting on subsystem $S$. Note that the partial trace condition is linear in $\rho_i$. Therefore, it can be naturally imported into \eqnref{eq:TrivialSdP} and still be formulated in terms of a SdP:
\begin{equation}
    \begin{array}{llr}
    \beta_{1}:=&\min_{\{\rho_i\}}& \sum_{i} \mathrm{Tr}[\rho_i \hat{H}_i]\\
    &\mathrm{s.t.}& \rho_{i} \succeq 0\\
    &&\mathrm{Tr}[\rho_i] = 1\\
    &&\mathrm{Tr}_{\mathrm{supp}(\rho_j)^c}[\rho_i] = \rho_{i \wedge j}.
    \end{array}
    \label{eq:SdPLevel1}
\end{equation}

Given that the sets of $\{\rho_i\}$ that satisfy the constraints of \eqnref{eq:SdPLevel1} also satisfy the constraints of \eqnref{eq:TrivialSdP}, we have $\beta_{\emptyset} \leq \beta_{1} \leq E_0$, by construction.

The certificates obtained from \eqnref{eq:SdPLevel1} can be further strengthened by adding virtual RDMs. For instance, even if $H$ is $2-$local, we might want to ask e.g. that the two-body RDMs acting on $Alice-Bob$ and $Bob-Charlie$ are such that they both come from a virtual three-body density matrix acting on $Alice-Bob-Charlie$. The latter is not strictly necessary in order to compute the energy, for $2-$body density matrices suffice, but this compatibility condition further restricts the set $\{\rho_i\}$, hence improving the bound. In mathematical jargon, this method is known as representing the feasible set as a projected spectrahedra \cite{ParriloBook2013}. Hence, instead of solely asking that $\rho_i$ and $\rho_j$ yield the same RDM on their intersection, now we might impose a stronger constraint, which is that $\rho_i$ and $\rho_j$ come from a valid density matrix $\rho_{i \vee j}$ defined on the union of their supports:
\begin{equation}
    \begin{array}{llr}
    \beta_{2}:=&\min_{\{\rho_{i\vee j}\}}& \sum_{i} \mathrm{Tr}[\rho_i \hat{H}_i]\\
    &\mathrm{s.t.}&\rho_{i \vee j} \succeq 0\\
    &&\mathrm{Tr}[\rho_{i \vee j}] = 1\\
    &&\mathrm{Tr}_{\mathrm{supp}(\rho_i)^c}[\rho_{i \vee j}] = \rho_{i}.
    \end{array}
    \label{eq:SdPLevel2}
\end{equation}
We observe that the constraints imposed in \eqnref{eq:SdPLevel2} automatically imply those of \eqnref{eq:SdPLevel1}, so we have omitted their writing, as they became redundant.

We also observe that, although now we have $\beta_{\emptyset} \leq \beta_{1} \leq \beta_2 \leq E_0$, the cost of solving \eqnref{eq:SdPLevel2} is substantially higher than that of \eqnref{eq:SdPLevel1}, because the SdP variables $\rho_{i \vee j}$ act on more qubits than $\rho_{i}$ and the cost of representing them grows exponentially in the number of qubits. Similarly, the relaxations \eqnref{eq:SdPLevel2} can be strengthened further by considering compatibility with more regions, yielding a chain of inequalities $\beta_{\emptyset} \leq \beta_{1} \leq \beta_2 \leq \ldots \leq  E_0$.

In \eqnref{eq:SdPLevel2} the compatibility constraints are enforced on all possible pairs $(i,j)$. However, not all the constraints are equally useful. In an extreme case, when $\mathrm{supp}(\rho_i) \cap \mathrm{supp}(\rho_j) = \emptyset$, adding the variable $\rho_{i \vee j}$ with its respective constraints makes no difference. Indeed, since $\mathrm{Tr}[\rho_i \hat{H}_i + \rho_j \hat{H}_j] = \mathrm{Tr}[(\rho_i\otimes \rho_j)(\hat{H}_i \otimes \mathbbm{1}_j + \mathbbm{1}_i \otimes \hat{H}_j)]$, the choice $\rho_{i \vee j } = \rho_i \otimes \rho_j$ is always possible, as it satisfies the rest of constraints, therefore not changing $\beta_2$. We remark this tensor product choice is possible because the supports do not intersect. However, if we define $\rho_{i \vee j}$ as a variable in \eqnref{eq:SdPLevel2}, we increase its computational complexity without improving the bound, thus yielding a worse certificate.

In \appref{app:SdPBasics} we give details on the basics of SdP and how to obtain mathematical proofs from their solutions.

\section{The constraint space}
\label{sec:constraintspace}
In this section we introduce the space of constraints for the relaxations and study its structure. This constraint space shall induce an underlying structure for the action space of the reinforcement learning agent in Section \ref{sec:constraintopt}. Following our running example, let us consider a set of $n$ qubits, labelled from $0$ to $n-1$, and denote $[n] = \{0, \ldots, n-1\}$. Let ${\cal P}([n]) = \{\emptyset, \{0\}, \{1\}, \ldots, \{n-1\}, \{0,1\}, \{0,2\}, \ldots, [n]\}$ denote the parts of $[n]$; \ie, the set of all subsets of $[n]$, thus containing $2^n$ elements.

Our first observation is that, to every subset $C \subseteq{\mathcal{P}}([n])$, we can associate a certificate in the following way: for each element $S\in C$, which corresponds to a subset of $[n]$, we consider the RDM acting on the qubits labelled by the elements in $S$, which we denote $\rho_S$. Let us denote $\Xi_C:= \{\rho_S\}_{S\in C}$ the collection of RDMs associated to $C$. By enforcing compatibility on their overlapping supports, we can define the SdP
\begin{equation}
    \displaystyle
    \label{eq:SdPCustom}
    \begin{array}{llrr}
    \beta_{C}:=&\min_{\Xi_C}& \sum_i \langle H_i \rangle&\\
    &\mathrm{s.t.}&\rho_{S} \succeq 0 & \forall S \in C\\
    &&\mathrm{Tr}[\rho_S] = 1&\\
    &&\mathrm{Tr}_{R^c}[\rho_S] = \mathrm{Tr}_{R^c}[\rho_{S'}]&\forall R \subseteq S \cap S',\quad S, S' \in C,
    \end{array}
\end{equation}
where the partial trace over the whole system is set to one by convention $\mathrm{Tr}_{[n]} [\rho]= 1$. We have written the objective function as $\sum_i \langle H_i \rangle$ for the following reasons: first, $C$ could be small enough so that there is no $S \in C$ such that $\mathrm{supp}(H_i) \subseteq S$. If this is the case, then we substitute $\langle H_i \rangle$ by the minimal eigenvalue of $H_i$, in the same spirit as the trivial relaxation \eqnref{eq:TrivialSdP}. Hence, if $C= \emptyset$, the cost function of \eqnref{eq:SdPCustom} amounts to the sum of the minimal eigenvalue of each $H_i$. Otherwise, if $\Xi_C$ contains a density matrix $\rho_i$ whose support contains the support of $H_i$, we simply compute $\langle H_i \rangle = \mathrm{Tr}[\rho_i H_i]$. Note that, in case that multiple density matrices from $\Xi_C$ could be used to compute $\langle H_i \rangle$, the last constraint of \eqnref{eq:SdPCustom} guarantees the result is well-defined; \ie independent of the choice $\rho_i \in \Xi_C$.
In practice, the last constraint of \eqnref{eq:SdPCustom} rarely needs to be imposed over all the subsets of the intersection, and it is enough to take $R=S\cap S'$ for all pairs $S,S' \in C$. In \appref{app:Pathological} we discuss these implications in a detailed way. Regardless of the constraint implementation of \eqnref{eq:SdPCustom}, a valid lower bound is yielded by the SdP.

Furthermore, given a set of constraints $C\subseteq {\cal P}([n])$, it is not necessary to define \eqnref{eq:SdPCustom} over all the variables contained in $\Xi_C$. If some $S\in C$ is contained in another $S' \in C$, such that $S\subseteq S'$, we can simply use $\rho_{S'}$, as it contains all the information on $\rho_S$. This choice is well-defined due to the constraints in \eqnref{eq:SdPCustom} and it naturally defines a simplification function $s: \Xi_C \mapsto s(\Xi_C)$, which allows to simplify the SdP by removing redundant variables.

One of the main motivations of this work is to optimize the quality of the lower bound within a limited computational budget. The asymptotic complexity of an SdP with $m$ variables of matrix size $n$ depends on the method that it is used to solve it. A rough estimate is $O(m^2n^2)$, but iteration costs of the algorithm are not factored in \cite{NavascuesPRA2009}. There exist interior-point methods which are faster than the ellipsoid method \cite{GroetschelBook1993}, \eg Alizadeh's algorithm runs in $\tilde{O}(\sqrt{m}(m+n^3)L)$ time, where $L$ is an input parameter and the $\tilde{O}$ notation is used to supress $\mathrm{polylog}(mn/\varepsilon)$ terms, where $\varepsilon$ is the required precision \cite{AlizadehSIAM1995, AroraFOCS2005}. In our case, we use the self-dual minimization method SeDuMi \cite{SeDuMi}, which has a complexity $\tilde{O}(m^2n^{5/2} + n^{7/2})$ for large-scale instances, although there are algorithms of $\tilde{O}(nm^3)$, suitable for small matrix sizes \cite{PeaucelleGuide2002}. Interestingly, quantum algorithms have been proposed to solve SdP \cite{BrandaoFOCS2017}, and machine learning methods have been studied to aid the SdP solver \cite{Krivachy2020}.

In light of the whole zoo of algorithms for SdP and their various complexities, it is clear the time complexity of an SdP instance is highly dependent on the solver used. Nevertheless, for our case study it is important that a given computational budget will determine a set of maximal $(m,n)$ that are allowed, which we estimate by effectively limiting the size and contents of $\Xi$ (cf. \secref{sec:benchmarking}).

\begin{figure}
    \centering
    \includegraphics[width=0.35\textwidth]{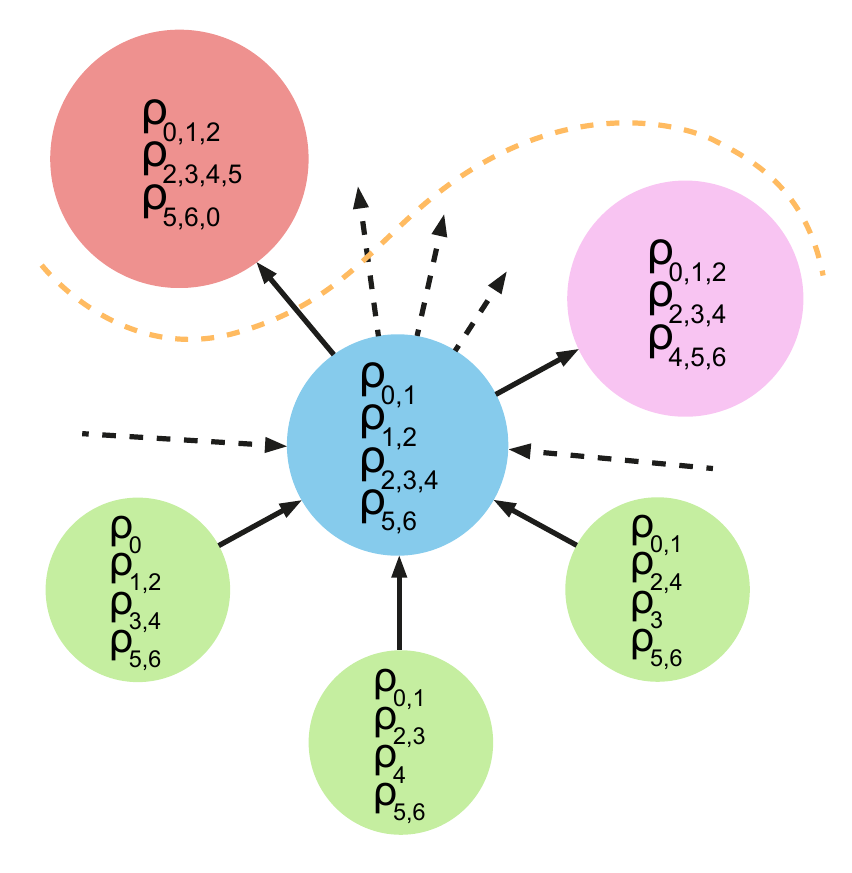}
    \caption{Poset structure of the constraint space. The different circles represent $\Xi_C$ for different $C\subseteq {\cal P}([n])$. The arrows represent the partial order relation $\preccurlyeq$ so that $\Xi_C \preccurlyeq \Xi_{C'}$ is represented from an arrow from $\Xi_C$ to $\Xi_{C'}$. Only the arrows relative to the central node are drawn. Dashed arrows indicate that there exist many more $\Xi_{C''}$ arriving/departing from the central node that are simply not drawn. The orange dashed line separates those $\Xi_C$ that fall into the allowed computational budget (green, blue and pink nodes) from those that are too expensive (red). Moving vertically up into the diagram provides better certificates, but at a higher cost. Since $\preccurlyeq$ is a partial order relation, some nodes (\eg the three at the bottom) are incomparable.}
    \label{fig:poset}
\end{figure}

The space of constraints forms a partially ordered set (poset) with respect to the following partial order relation. Given ${C, C'} \in {\mathcal{P}([n])}$, we say $C \preccurlyeq C'$ if, and only if, for each $S \in C$ there exists a $S' \in C'$ such that $S \subseteq S'$. The motivation of the partial order relation $\preccurlyeq$ is that $C \preccurlyeq C'$ implies $\beta_C \leq \beta_{C'}$ by construction: every density matrix in $\Xi_C$ can be obtained by tracing out some elements of another density matrix in $\Xi_{C'}$, and the constraints in \eqnref{eq:SdPCustom} enforce mutual compatibility among all the elements in $\Xi_C$ and $\Xi_{C'}$. In \figref{fig:poset} we illustrate such structure, which motivates the agent definition in \secref{sec:constraintopt}.

\section{Constraint optimization}
\label{sec:constraintopt}

In this section we discuss a method to achieve the best trade-off between the computational cost and the quality of a certificate by exploring the constraint space described in \secref{sec:constraintspace}. Hence, we face a constrained optimization problem over the constraint space, subject to the computational budget. Due to the high amount of structure in this extensive combinatorial space, we propose to use Reinforcement Learning (RL)~\cite{SuttonBook2018} with function approximation, which, with our proposed framework, naturally prefers lower cost solutions and is able to optimize its exploration strategy based on previous experiences. In such spaces, experience in one region may be useful in others, \eg in periodic systems, actions in one domain should be identical to actions in another, which further allows for easy transfer of learning without explicit analysis of the model parameters (see~\secref{sec:transfer}).

\begin{figure}
    \centering
    \begin{subfigure}{0.5\textwidth}
        \includegraphics[trim=25 40 25 40,clip,width=\textwidth]{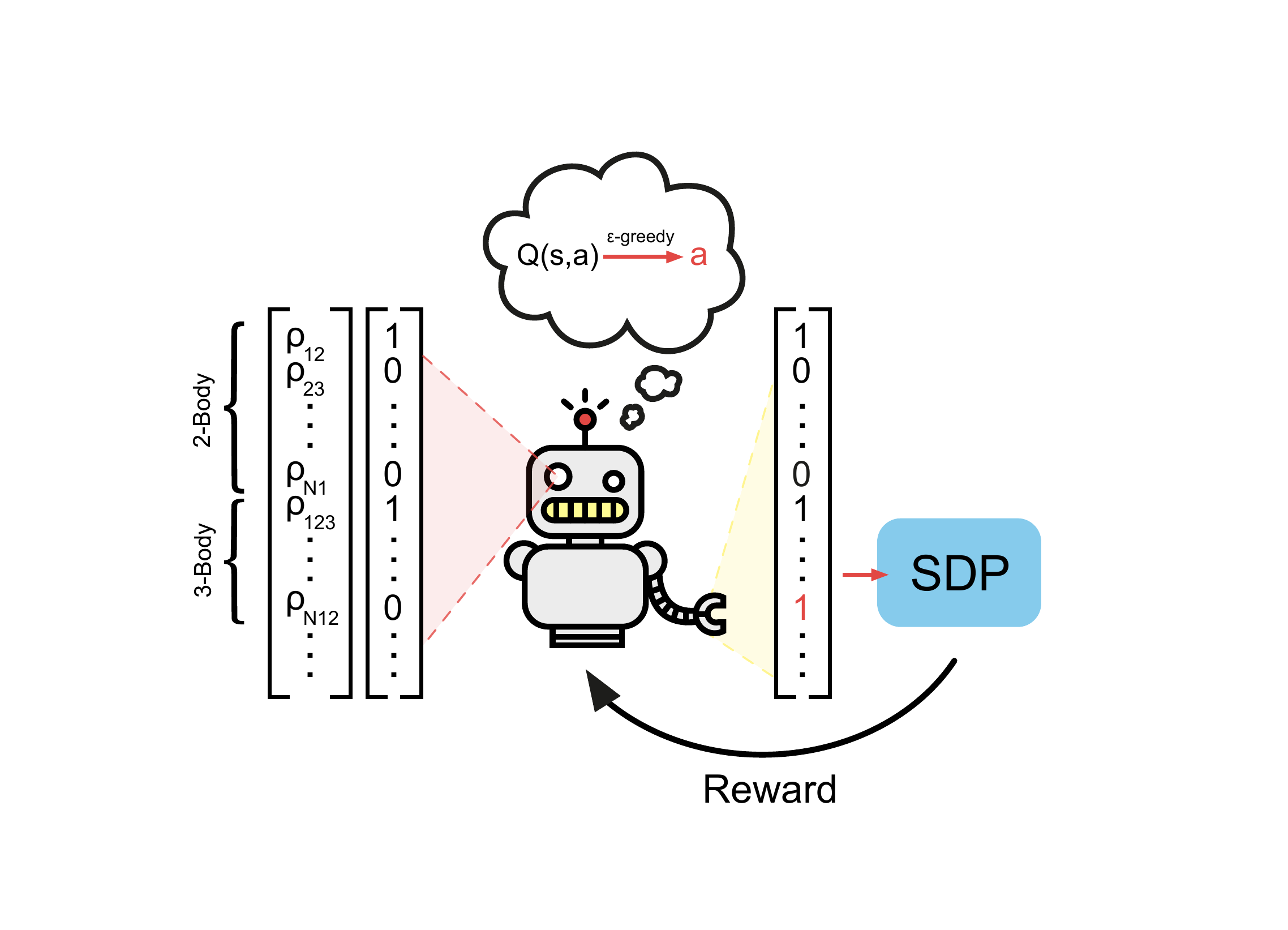}
        \caption{Reinforcement learning framework diagram.}
		\label{fig:RL_framework}
    \end{subfigure}%
    \begin{subfigure}{0.4\textwidth}
    \centering
        \includegraphics[width=\textwidth]{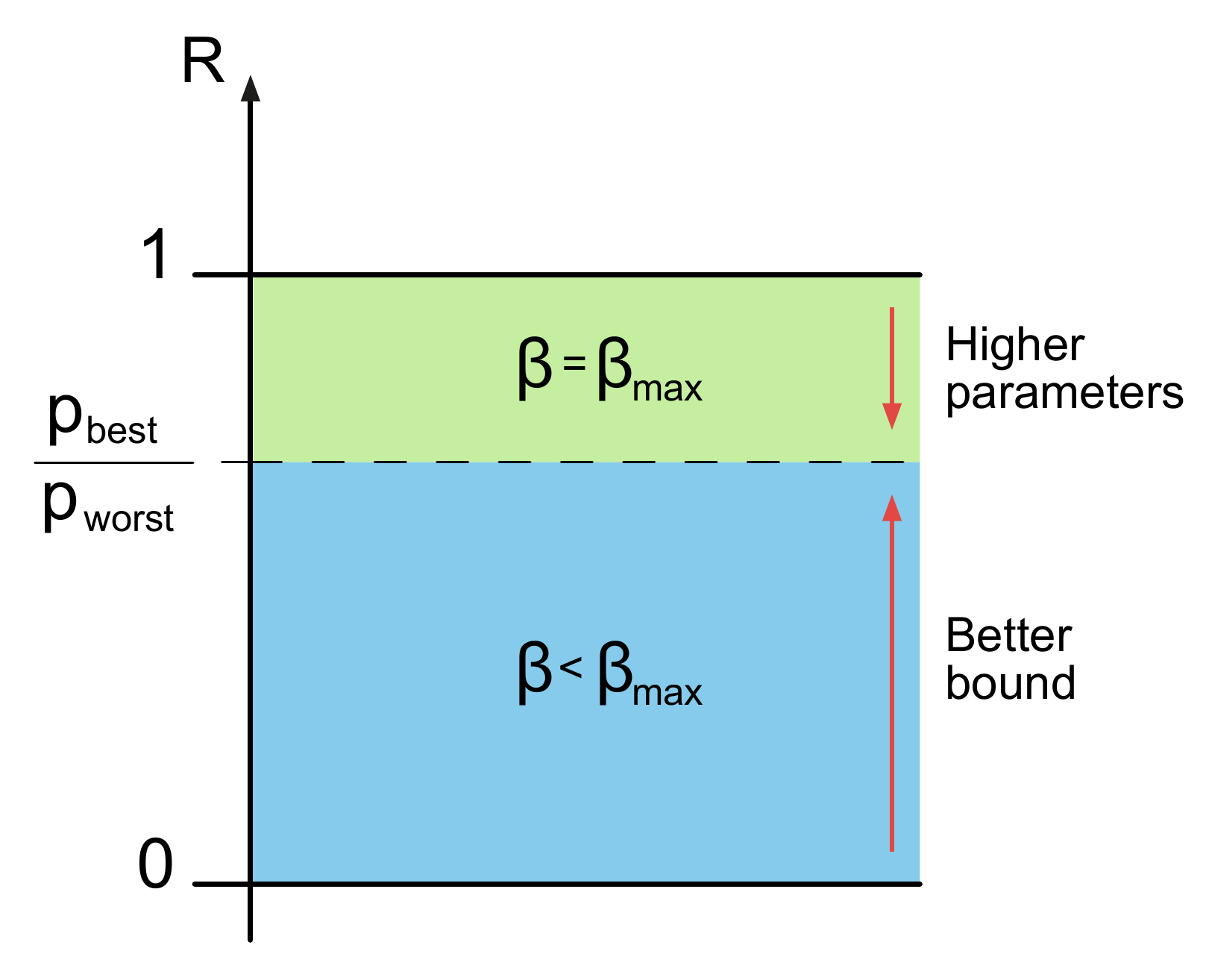}
        \caption{Reward function diagram}
        \label{fig:reward}
    \end{subfigure}
    \caption{(a) Schematic representation of the reinforcement learning framework. First, the agent observes the state: a one-hot encoding of the active constraints. Given the observation, it estimates the Q-values associated to the possible actions with the deep Q-network. Then, it decides which action to take according to an $\epsilon$-greedy policy, which results into a new set of constraints, bringing the agent to a new state. Finally, the black box solves the SdP associated to the new state, providing the agent with a reward, whose parts from~\eqnref{eq:reward} are illustrated in (b).}
\end{figure}

To this end, we frame the optimization problem as a Markov decision process (MDP). The MDP is defined through a state space, an action space, a transition function between states given an action and a reward function, which associates a value to each state-action-state tuple. All the parts are detailed below. A learning agent, as the learning program is called in RL terminology, explores the constraint space with the goal to find the set of constraints $C^* \subseteq {\cal P}([n])$ that provides the best possible certificate within a limited computational budget, while using the least amount of resources. In algorithmic terms, we distinguish two main independent parts:

\renewcommand{\theenumi}{\roman{enumi}}
\begin{enumerate}
    \item A black box, acting as reward function. It takes a set of constraints $C$ as input, computes $\beta_C$ by solving the associated SdP (\eqnref{eq:SdPCustom}) and outputs a reward, which depends on the quality of the resulting bound and its computational cost.
    \item A learning agent capable of generating sets of constraints and inputting them into the black box (i). The agent can choose to strengthen or loosen the constraints, effectively exploring the constraint space with its actions. In doing so, the agent obtains different rewards that guide it towards finding the optimal relaxation. Note that the agent is completely agnostic about the actual physical problem at hand. 
\end{enumerate}
We aim to understand up to which extent such a fully automated approach may help in studying physical systems. In the following, we connect the MDP components to our running example. See \figref{fig:RL_framework} for a schematic depiction.

\textbf{State space --}
The state space corresponds to the constraint space introduced in \secref{sec:constraintspace}, in which each state is a specification of constraints $C\subseteq\mathcal{P}([n])$ and it is bound by the computational budget, as illustrated in~\figref{fig:poset}. 
We represent the states by one-hot encoding of the active constraints $S\in C$: considering a set of $2^n$-dimensional canonical vectors with only a non-zero unit element, each representing an element $S\in\mathcal{P}([n])$, a state vector is the sum of the vectors that encode the components $S\in C$. Equivalently, it identifies the set $\Xi_C = \{\rho_S\}_{S\in C}$ of RDMs that enter as variables in \eqnref{eq:SdPCustom}. As shown in the leftmost part of \figref{fig:RL_framework}, the RDMs $\rho_S$ are ordered according to their dimension in the state vector. Out of the $2^n$ possible variables, we need only consider $\mathrm{poly}(n)$ of them, effectively reducing the state vector size: we can ignore the $1$-body constraints as well as those $\rho_S$ whose sole contribution to the cost of solving the associated SdP would exceed the computational budget. With a  computational budget $B$, this leaves $n^{O(\log(B))}$ available RDMs to construct the certificate. If no $S\in C$ is such that $i \in S$ the $1$-body constraint corresponding to $\rho_{\{i\}}$ is added by default. Therefore, the smallest set of constraints that we allow for is $C=\{\{0\},\dots,\{n-1\}\}$, represented by a state vector of zeros, and we take it as the initial state of the MDP. 

\textbf{Actions --} An action $a$ consists of either adding or removing a constraint, driving the agent from one state to another. In practice, actions flip bits in the state vector corresponding to the encoded constraints. The agent is free to add a constraint of any size, as long as the cost associated to the resulting set is within the computational budget. For instance, the agent can start by adding a 4-body constraint, e.g. $\rho_{0123}$, to the initial state. In contrast, removing a constraint has a different effect. In order to keep the state space exploration consistent, removing a constraint splits it into its most immediate components of a lower degree. For instance, in 1D, removing $\rho_{0123}$ would result into $\rho_{012}$ and $\rho_{123}$. Note that a valid action always corresponds to an arrow (in both directions) in the poset depicted in~\figref{fig:poset}.

\textbf{Transition function --} The transition function is a simple deterministic function implicitly defined above: $T(C| a, C')$ is a Kronecker delta, attaining unit value if the constraint configuration $C$ is reached by adding or removing the constraint specified by the action $a$ from the  set of constraints $C'$.

\textbf{Reward --} The reward function is defined to match the overall optimization goal, provided that the learning agent aims to maximize the obtained reward. The reward associated to a state $C$ depends on: 1) the energy bound $\beta_C$, obtained solving its associated SdP, and 2) its computational cost. In practice, we take the amount of free parameters in the SdP \eqnref{eq:SdPCustom}, which we denote by $p$, as a representation of the computational cost. Note that, given an initial, unconstrained, optimization problem, we have no prior knowledge about the optimal $\beta$ and $p$. Therefore, in order to compute the reward associated to a given state, we rely on a set of references that are updated as the constraint space is explored. More precisely, we keep track of the best and worst bounds obtained, $\beta_{\max}$ and $\beta_{\min}$ respectively, and the best and worst set of parameters with which the best bound so far $\beta_{\max}$ has been observed, denoted $p_{\text{best}}$ and $p_{\text{worst}}$ respectively. The reward associated to a state is computed by comparing the actual $\beta$ and $p$ to the reference values as
\begin{equation}
R(\beta, p) = \frac{p_{\text{best}}}{p_{\text{worst}}} \cdot
\begin{cases}
\frac{p_{\text{worst}}}{p} & \text{if $\beta=\beta_{\max}$} \\
\left(\frac{\beta - \beta_{\min}}{\beta_{\max} - \beta_{\min}}\right)^d & \text{otherwise},
\end{cases}
\label{eq:reward}
\end{equation}
where $d$ is a fixed exponent that controls the shape of the line $(\beta - \beta_{\min})/(\beta_{\max} - \beta_{\min})$. Such exponent is introduced in order to provide better discrimination 
depending on how close to each other are the different bounds obtained for different $C$.
Notice that $p_{\text{worst}} \geq p_{\text{best}}$ and, therefore, $p_{\text{worst}}/p \geq 1$. Thus, the prefactor $p_{\text{best}}/p_{\text{worst}} \leq 1$ ensures that $R(\beta, p) \in [0, 1], \ \forall \ \beta,p$. \figref{fig:reward} shows a schematic of the reward function. In summary, the reward function mainly focuses on the resulting bound $\beta$, unless various states provide the maximum possible bound $\beta_{\text{max}}$. In this case, those with higher computational costs are penalized.

\textbf{The agent --} Within the proposed framework, the constrained optimization can be solved through various methods. As mentioned before, we propose to use RL with function approximation. The learning program or agent specifies the policy by which actions are taken, with the ultimate goal of maximizing the obtained reward. More precisely, we use double deep Q-learning~\cite{PhDWatkins,MnihNature2015,vanHasseltAAAI2016} with an $\epsilon$-greedy policy $\pi$. At each state $C$, the agent estimates the Q-values $Q^\pi(a,C)$ of each possible action $a$, a measure of the expected rewards associated to taking each action and then following the policy $\pi$. The $\epsilon$-greedy policy considers that the actions are taken according to
\begin{equation}
 \pi(C) = 
\begin{cases}
    \argmax_a Q^\pi(a,C), &\text{with probability  } (1-\epsilon)\\
    \text{uniform random } a, &\text{with probability  } \epsilon.
\end{cases}
\label{eq:policy}
\end{equation}

\figref{fig:RL_framework} shows a schematic representation of the whole process. In \secref{sec:results} we show that such approach leads to solutions faster compared to other classical optimization methods and, sometimes, it is even able to find the optimal solution where the other methods fail.

\subsection{Application to the Heisenberg XX model}
\label{sec:results}

Following our running example of finding a lower bound to the ground state energy of quantum local Hamiltonians, we focus on a paradigmatic condensed matter model: the anti-ferromagnetic 1D quantum Heisenberg XX model~\cite{LiebAnnPhys1961}, described by the Hamiltonian
\begin{equation}
\label{eq:XX}
    H=\sum_{i=0}^{n-1} J_{i}(\sigma_{i}^x \sigma_{i+1}^x + \sigma_{i}^y \sigma_{i+1}^y) + \sum_{i=0}^{n-1}B_{i}\sigma_{i}^z,
\end{equation}
where $\sigma^\alpha, \ \alpha=x, y, z$, are the Pauli matrices, $J_i$ is the antiferromagnetic exchange interaction between spins and $B_i$ is the strength of the external magnetic field. We consider periodic boundary conditions, such that $\sigma_{n}^\alpha=\sigma_{0}^\alpha$. In the homogeneous case, \ie $J_i=J, B_i=B \ \forall i$, the model presents a quantum phase transition at $B=2J$~\cite{SachdevBook2009} between an antiferromagnetic and a paramagnetic phase, in which the entanglement vanishes~\cite{WangPRA2001, WangPRA2002, PasqualeEPJST2008}. We will hence refer to these phases as entangled and unentangled, respectively. Although the 1D XX model \eqnref{eq:XX} is efficiently solvable via the Jordan-Wigner transformation \cite{JordanWigner1928}, corresponding to a quadratic fermionic Hamiltonian \cite{NielsenNotes2005, BanulsPRA2007, TuraPRX2017}, the agent is oblivious to such information. We emphasize that the points in the search space have no semantics to the agent, which, moreover, is not provided with any information about the Hamiltonian in any explicit way. This guarantees that our approach is as generally applicable as possible.

\subsubsection{Results}
\label{sec:opt_results}
We present the results on the application of the RL method to the homogeneous version of the aforementioned Hamiltonian. The maximum budget considered for this example allows for the allocation of half of the possible 3-body constraints and all the 2-body ones. Given the budget, we proceed with finding the best approximation to the ground state in the whole phase diagram of the Hamiltonian. 

\begin{figure}
    \centering
    \includegraphics[width=\textwidth]{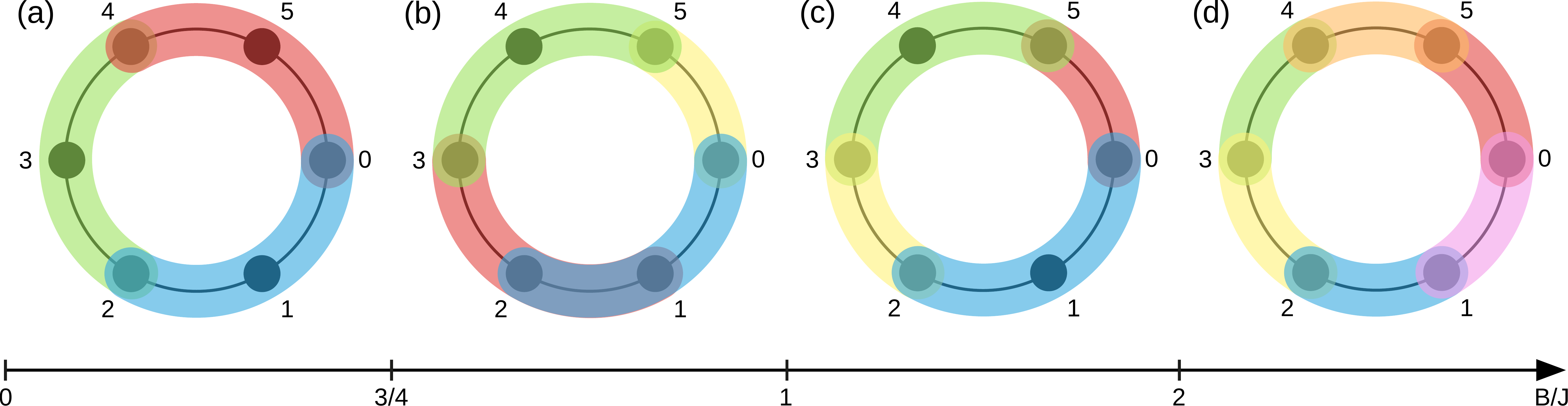}
    \caption{Illustrative representation of the RDMs, whose support is depicted in different colors, considered in the SdP optimization to obtain the best energy bound of the homogeneous Heisenberg XX model,~\eqnref{eq:XX}. Compatibility constraints are imposed over the overlapping areas. The results are obtained with a budget that allows for the allocation of up to half of the 3-body RDMs and $n=6$. For this case, the RL algorithm finds four different optimal solutions. Interestingly, the entangled phase ($B/J<2$) shows two intermediate solutions: (c) and (b), before the expected set of constraints (a) at $B/J<3/4$.}
    \label{fig:toy_examples}
\end{figure}

 \textbf{Unentangled phase} $\mathbf{B/J\geq2}$ \textbf{--} In the unentangled phase, the ground state can be perfectly described by the set of independent 1-body RDMs. Therefore, we would expect the optimal set of constraints to be the minimum that the agent can consider $C=\{\{0\},\dots,\{n-1\}\}$. Nevertheless, this is only true in the extreme case of $J=0$. In a general scenario, with $0<2J\leq B$, the optimal solution is made out of 2-body constraints, as shown in~\figref{fig:toy_examples} diagram (d). This is to provide support for the 2-body terms of the local Hamiltonian. Recall that, in our implementation, whenever a term $H_i$ of the Hamiltonian is not supported by the set of RDMs $\Xi_C=\left\{\rho_S\right\}_{S\in C}$, we take $\langle H_i \rangle$ to be its minimal eigenvalue $\min(\sigma(H_i)) = -J$. With 2-body constraints, the resulting RDMs are rank-$1$ projectors, which correspond to pure states such that $\langle H_i \rangle = 0$ for the 2-body terms, thus yielding a better energy bound. Increasing the size of the constraints any further does not improve the energy bound at all.

\textbf{Entangled phase} $\mathbf{B/J<2}$ \textbf{--} In the case of the entangled ground state, its exact energy can only be obtained by considering the system as a whole, corresponding to $C=\{[n]\}$. Therefore, the agent can only provide the best possible approximation to the exact energy within the allowed computational budget. Just like in the previous case, it may seem reasonable to expect the optimal set of constraints to be unique for the whole phase. Nevertheless, the agent finds three separate regimes as depicted in~\figref{fig:toy_examples}:
\begin{itemize}
    \item Close to the phase transition, the best certificate is obtained by alternating 2-body and 3-body constraints, as shown in~\figref{fig:toy_examples} diagram (c). This solution has a lower complexity than (a) and (b), but it provides a higher energy bound.
    \item In an intermediate regime, as shown in~\figref{fig:toy_examples} diagram (b), the best possible certificate is obtained combining the overlap of some of the largest possible constraints with the inclusion of a smaller constraints.
    \item Deep into the phase, as shown in~\figref{fig:toy_examples} diagram (a), the best possible certificate is obtained by evenly distributing all the largest possible constraints throughout the system. A priori, we would expect this to be the optimal solution throughout the whole phase.
\end{itemize}

Note that, in the entangled phase, the two intermediate optimal configurations (b) and (c) provide better bounds than the set of constraints (a) in~\figref{fig:toy_examples}, even with (c) yielding simpler certificates. This simple scenario shows that evaluating the quality of a relaxation beforehand is not a trivial task and it becomes even less straightforward when considering different kinds of Hamiltonians and budgets. Additionally, a budget that allows the allocation of several 3-body RDMs, may also allow for the allocation of some 4-body constraints, which are also taken into account in the optimization. For instance, with $n=10$, the agent could introduce a single 4-body constraint. However, the solution found by the agent shows that it is better to combine 3-body and 2-body RDMs rather than using such a limited amount of 4-body ones.

In~\figref{fig:toy_examples} we show the solution of a small system of $n=6$ sites for illustrative purposes. In larger systems, we observe that the same optimal patterns remain consistent, suggesting that the qualitative solutions obtained in small systems can be used at larger ones with similar properties. In \appref{app:SolsLargerN} we provide further details about the quality of the obtained certificates throughout the phase space and show that the optimal sets of constraints do remain optimal across different system sizes. While a thorough characterization of the Hamiltonian in terms of its optimal SdP constraints is of great interest, it falls out of the scope of the work. Already in such a simple scenario, the agent is able to find a rich set of intermediate solutions, which may, at first glance, seem counter-intuitive. The solutions are, nevertheless, closely related to the actual entanglement structure of the ground state of the system~\cite{WangPRA2002}. This shows that the agent is able to capture physical properties of the system, even when various possible solutions are very close in terms of cost and quality.

As a final remark, see that, the ground state of the unentangled phase is a product state, meaning that the exact solution lies within the budget with which the agent is provided. In contrast, in the entangled one, the ground state can only be exactly described by its full density matrix, meaning that the exact solution falls outside of the budget. With the framework we here present, when the agent is far from using the whole budget, it may be seen as a strong indication that the provided result is exact (cf. Section \ref{sec:Particularcases}). 

\subsubsection{Benchmarking}
\label{sec:benchmarking}

As briefly introduced at the beginning of~\secref{sec:constraintopt}, the proposed framework allows for the straightforward application of several optimization algorithms, besides RL. In this section, in order to evaluate the quality of the RL results, we use two informative points of reference: breadth first search (BFS)~\cite{CormenBook2009} and Monte Carlo (MC) optimization~\cite{KirkpatrickScience1983}. To the best of our knowledge, this is the first time such kind of optimization is performed. Thus, not having a pre-defined benchmark, we establish the first steps. 

For the comparison, we consider an inhomogeneous version of the XX Heisenberg model~\eqnref{eq:XX} in which we keep a constant magnetic field $B_i=B=1$ and tune the interaction strength $J_i=i\mod 3$. This provides us with isolated groups of three interacting sites. Note that, depending on the system size, there may be exclusively triplets, triplets and an isolated site or triplets and a pair. Such model allows us to find out the optimal set of constraints beforehand, which lets us compare the performance of the optimization algorithms with respect to the actual optimal solution.

As a figure of merit to evaluate the algorithm's performance, at each time-step we compute the reward of the given state, as in~\eqnref{eq:reward}, with full knowledge of $\beta_{\max}, \beta_{\min}, p_{\text{best}}, p_{\text{worst}}$. This provides a measure of closeness to the optimal configuration, obtaining reward $1$ for the optimal state.   

Note that the algorithms have different ways to explore the state-space. Hence, in order to perform a fair comparison of the progress towards the optimal set of constraints, we do not take into account repeated visits to the states. Contrary to the the BFS, both the RL and the MC agents can go back and forth revisiting the same states several times. Given that the main computational cost comes from solving the associated SdP to each state, we keep a memory of the solutions already obtained throughout the path. Hence, we consider that revisiting a state implies a negligible computational cost. 

\begin{figure}
    \begin{subfigure}{.5\textwidth}
      \centering
      \includegraphics[width=\textwidth]{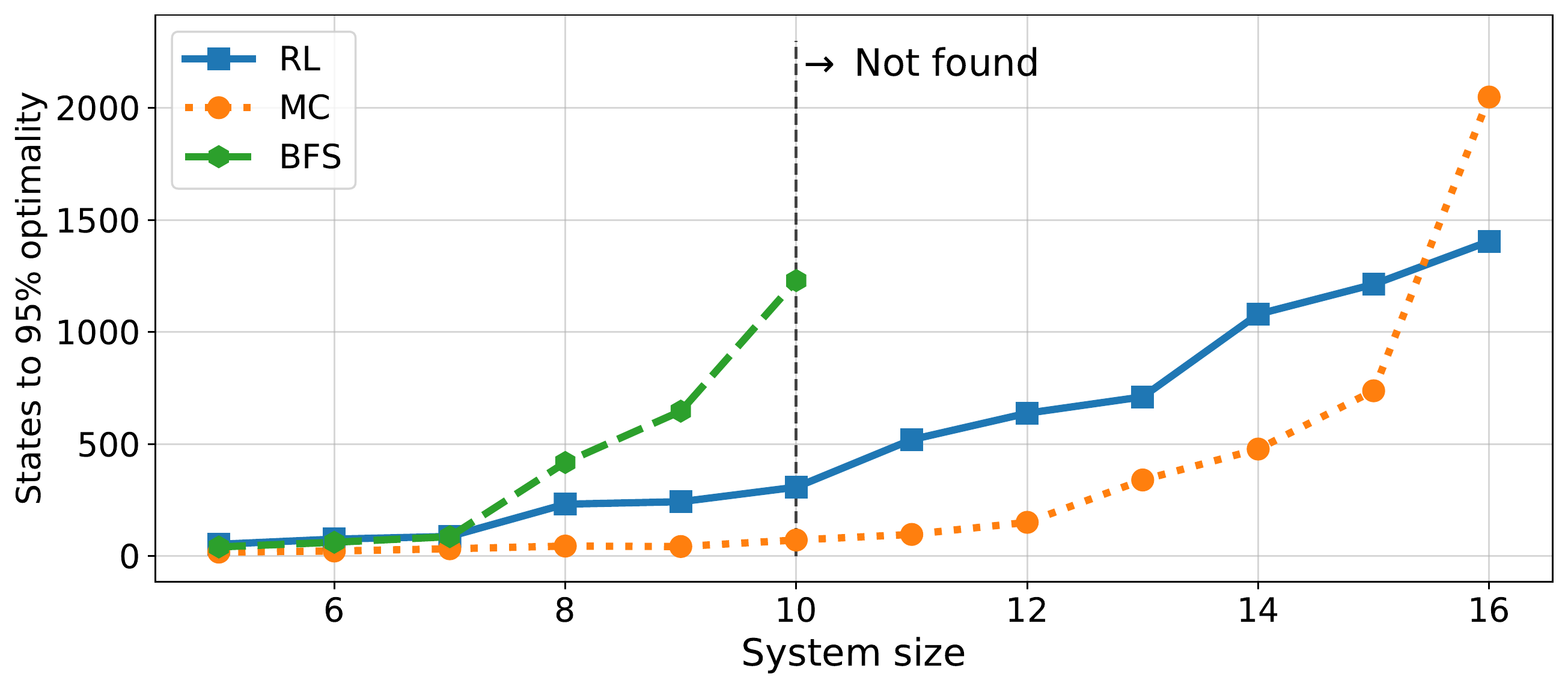}
      \caption{Benchmark over systems with a budget equivalent to half of the possible 3-body constraints.}
      \label{fig:bench_half}
    \end{subfigure}%
	\begin{subfigure}{.5\textwidth}
      \centering
      \includegraphics[width=\textwidth]{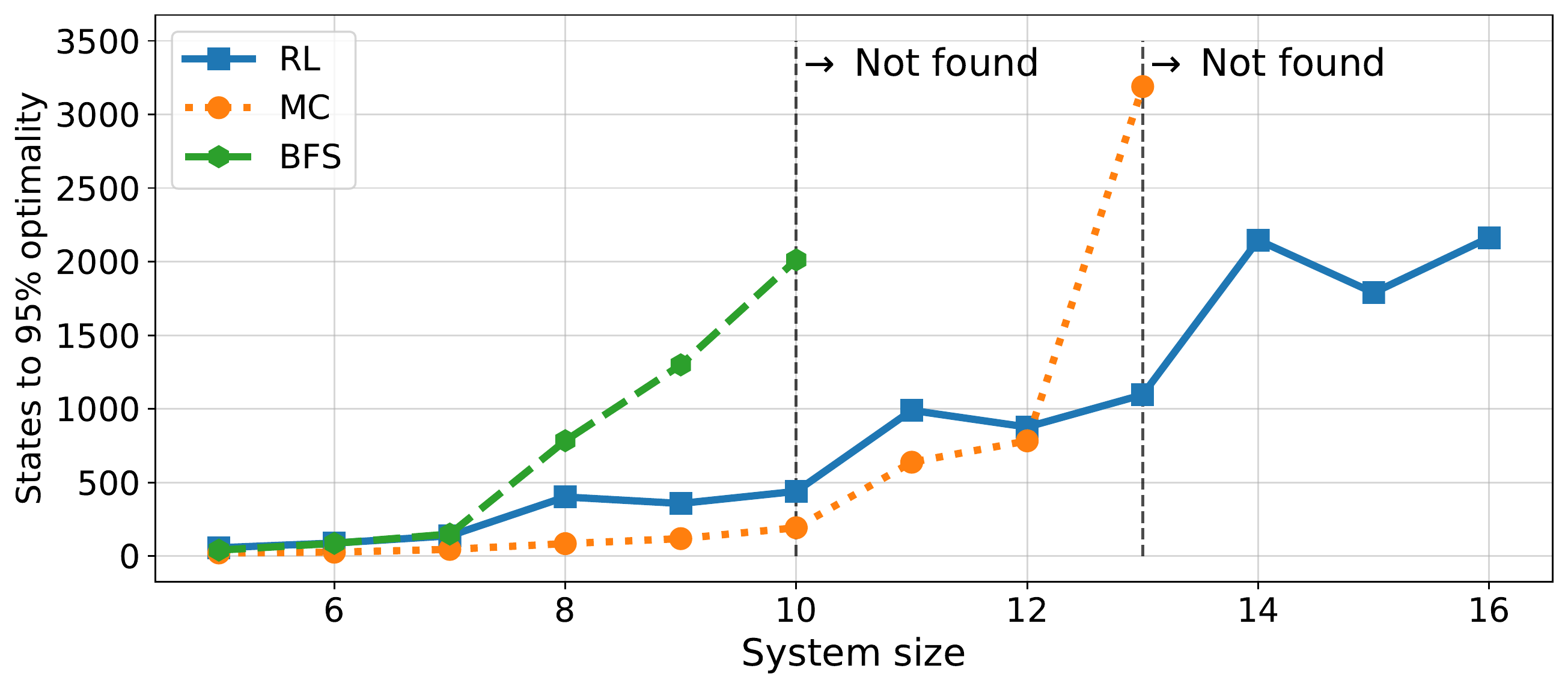}
      \caption{Benchmark over systems with a budget equivalent to all of the possible 3-body constraints.}
      \label{fig:bench_all}
    \end{subfigure}%
    \caption{Benchmark of the performance of the three optimization algorithms: Breadth first search (BFS), Monte Carlo (MC) and deep reinforcement learning (RL). The algorithms are evaluated in two scenarios: allowing up to (a) half of all the 3-body constraints and (b) all the 3-body constraints. The dashed vertical lines indicate the system size beyond which the overlapping algorithm was unable to find the optimal state in less than 4000 visited states.}
    \label{fig:benchmark}
\end{figure}

Consequently, we evaluate the overall performance by keeping track of the best obtained reward for every new visited state. In~\figref{fig:benchmark}, we depict the amount of new states visited by fifty agents before, on average, they reach a reward of $0.95$. The process is repeated for several system sizes, with which the constraint space increases exponentially. The hyper-parameter tuning for the RL and MC optimizations are performed at a system size of $n=10$ and kept throughout the whole process (see \appref{app:Hyperparams}). 

First, we benchmark the agent performance providing them with a small budget, which allows the agents to allocate only half of the available 3-body constraints. The results are depicted in~\figref{fig:bench_half}. For small systems, there are no substantial differences in performance, given that the state space is reduced. Already at $N=11$, the BFS is not able to find the optimal bound within a reasonable time. While the MC optimization provides better results for small systems, it is out-performed by the RL agent at $N=16$. We hypothesize that, at this size, the overhead of learning is overcome by the increasing complexity of the state space.

In order to test this hypothesis, we conduct the same experiment with a larger computational budget that allows the agents to allocate all the 3-body constraints. With this, for the same system sizes, the agents encounter significantly larger constraint-spaces. The results are shown in~\figref{fig:bench_all}. In this case, the differences between the MC and RL optimizations are relatively smaller for smaller systems and the RL agents outperform the MC optimization earlier on. This means that, for large state spaces, the learning cost involved in the RL optimization pays off, making it better than following a simple MC heuristic. In addition, unlike the RL, the MC shows a strong dependency on a proper hyper-parameterization, \eg choosing an appropriate inverse temperature, provided that, as soon as the parameters are not optimised for the specific problem, the performance is dramatically affected. Proper parameter tuning is, in itself, a computationally costly task, given the constraint-space size. The RL scheme, being quite resilient to its hyper-parametrization, provides a significant advantage in this sense, allowing us to tune it in reduced systems.

\subsubsection{Transfer learning}
\label{sec:transfer}
An interesting feature of the proposed framework is that none of its parts require prior information about the actual problem. This suggests the possibility of exploring a given constraint optimization and its underlying system in a completely autonomous way. One way to take advantage of this feature is by performing transfer learning (TL)~\cite{TaylorJMLR2009}. In order to do so, we start by training an agent to solve a system under the action of a Hamiltonian. Then, we leverage the experience obtained by the agent in the initial task using it as initial condition to solve a new problem with a similar Hamiltonian.  

\begin{figure}
    \centering
    \includegraphics[width=0.8\textwidth]{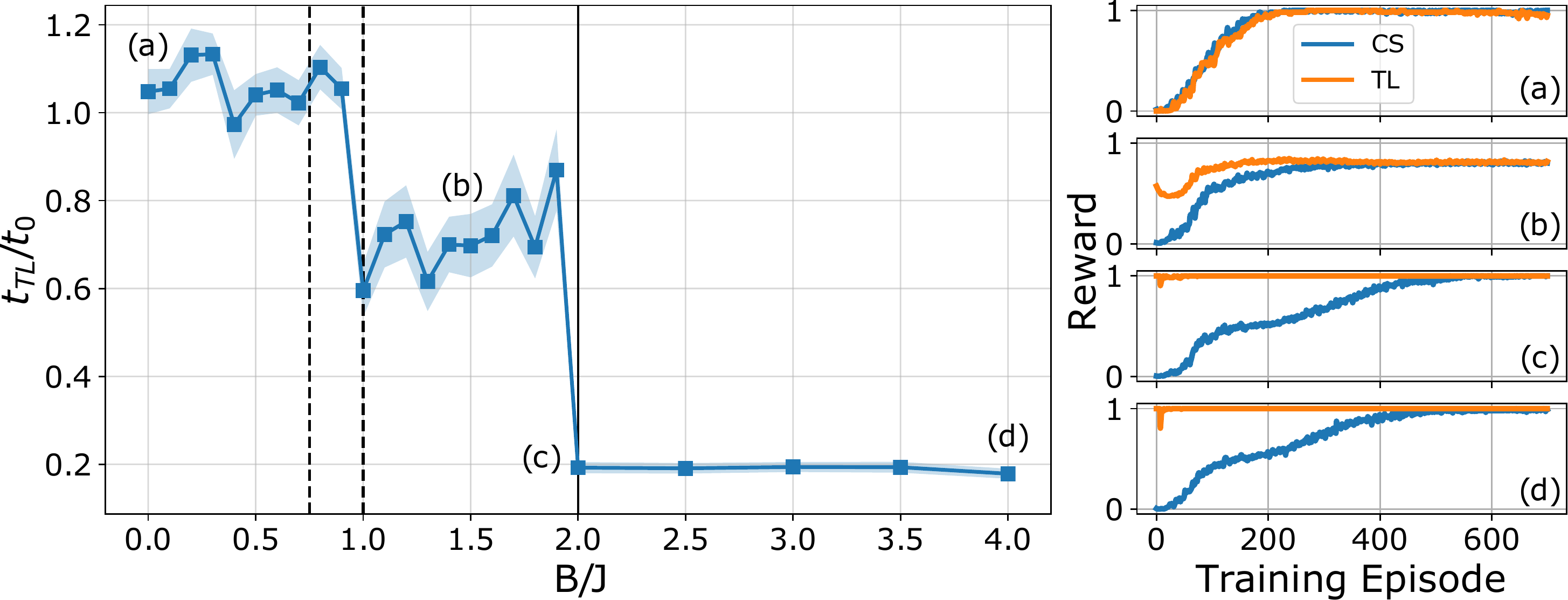}
    \caption{Transfer learning results for an ensemble of fifty independent agents. (Left) Convergence time ratio between transfer learning and cold start as function of the parameter $B/J$ (recall~\eqnref{eq:XX}). The pre-trained models are taken from $B/J=5$ and they are used as starting point in the optimization for different values of the parameter. The vertical lines indicate qualitative changes in the optimal solution. (Right) Reward obtained at the final state of an evaluation episode after each training episode with transfer learning (TL) and cold start (CS).}
    \label{fig:TL}
\end{figure}

We consider an homogeneous version of the Heisenberg XX model~\eqnref{eq:XX}. As commented before, this Hamiltonian shows a quantum phase transition at $B/J=2$, but also shows three different solutions in the entangled phase ($B/J<2$). An agent is trained to solve the constraint optimization deep in one phase, with $B/J=5$. Then, we use such agent to find the optimal solution for the rest of the phase space. In~\figref{fig:TL} we show the ratio between the time it takes the algorithm to converge with TL $t_{TL}$ and the time it takes with a cold start $t_0$, \ie a training starting from scratch. Hence, with $t_{TL}/t_0<1$ there is favorable TL and with $t_{TL}/t_0>1$ there is negative transfer. The convergence time is obtained averaging the results of training fifty independent agents, shown on the right panel of~\figref{fig:TL} (see also \cite{DawidNJP2020}).

We observe that TL in the same phase is quite favorable. Indeed, for this particular problem, the optimal set of constraints is the same across the whole phase, including the critical point (cases (d) and (c), respectively). When applied across phases, the advantage of TL diminishes sharply. Close to the phase transition (case (b)), there appears a local minimum in which some agents get stuck and, under the given conditions, it takes them hundreds of training episodes to correct it. In this regime, the TL still provides an advantage regarding convergence, although it does not help avoiding the sub-optimal configuration. Deep into the opposite phase (case (a)), even though TL barely affects the performance, as $t_{TL}/t_0\simeq1$, it has a slightly negative impact.

The vertical lines of \figref{fig:TL} show the phase transition (solid) and the intermediate points in which the optimal set of constraints changes (dashed). As shown, the loss of a convergence advantage from TL can  be indicative of changes in the ground state of the system. Hence, this approach can thus be used to infer the properties of the physical system in a completely unsupervised way, by exploiting the failure of the method such as in~\cite{NieuwenburgNatPhys2017, KottmannPRL2020}. 

\section{Particular cases}
\label{sec:Particularcases}
Here we analyze some cases where the local Hamiltonian \eqnref{eq:localH} enjoys desirable properties that make a certificate easier to obtain.
\begin{itemize}
    \item If $H$ is a frustration-free Hamiltonian, its lowest energy eigenstate coincides with a lowest energy state of each of the individual terms $H_i$. In other words, global ground states correspond to local ground states. In this case, let $\ket{\psi}$ be the ground state of $H$. It is also a ground state of every $H_i$ and it defines a set of RDMs $\rho_i = \mathrm{Tr}_{\mathrm{supp}(H_i)^c}\ket{\psi}\bra{\psi}$. Note that frustation-freeness guarantees that the contribution of each term equals its algebraic minimum $\mathrm{Tr}[\rho_i H_i] = \min \sigma(H_i)$. Hence, the minimal set of constraints $C_\emptyset$ (cf. Eq. \ref{eq:TrivialSdP}) already reproduces the ground state energy: on the one hand, given a term $H_i$, there is no $\rho_i$ that yields a smaller value than $\min \sigma(H_i)$. On the other hand, the set of RDMs that correspond to the actual ground state satisfy this condition. This implies that strengthening the constraints in \eqnref{eq:SdPCustom} to any $C \succcurlyeq C_\emptyset$ will be of no effect in increasing $\beta_C$.

    A couple of comments are in order:
    \begin{itemize}
        \item Obtaining a set of constraints $\Xi_C$ which recovers an exact lower bound $\beta_C = E_0$ does not automatically imply that we can recover the ground state configuration, even if the problem is fully classical. For instance, even if $H$ corresponds to a classical $3$-SAT problem: $H$ can be written in the computational basis as a sum of projectors $\Pi_i$ that act non-trivially on $3$ variables $x_{i_1}$, $x_{i_2}$ and $x_{i_3}$. Since $\Pi_i \succeq 0$ and there exists a satisfiable instance, we obtain $\beta_C = 0$ for any relaxation. By inspecting the values of the $\rho_i$ that the SdP \eqnref{eq:SdPCustom} outputs, it does not need to be the case that $\rho_i$ is a rank-$1$ projector onto the solution state $\ket{x_{i_1}x_{i_2}x_{i_3}}$ and thus directly interpretable as part of the solution to $3-$SAT.
        \item Frustration-free Hamiltonians constitute an important class of models. All short-range, gapped, Hamiltonians can be well-approximated by frustration-free ones by increasing their locality to be $O(\log(n))$ \cite{HastingsPRB2006}. Frustration-free Hamiltonians comprise notable models, both commuting and anticommuting:
        On the one hand, frustration-free, commuting models include the toric code \cite{KitaevAnnPhys2003, KitaevAnnPhys2006}, Levin-Wen models \cite{LevinPRB2005} and quantum error correcting codes \cite{Gottesman2009}. Importantly, graph states \cite{HeinVarenna2006} or, more generally, stabilizer states such as the cluster state \cite{BriegelPRL2001} are included in this class. Graph states can be approximated as ground states of two-body  Hamiltonians \cite{DarmawanNJP2014}, although it has been shown for spin-$1/2$ that this approximation cannot be made exact (ground states of frustration-free $2$-local qubit Hamiltonians are unentangled \cite{Bravyi2006, ChenPRA2011b}), even if we drop the frustration-freeness condition \cite{NielsenRepMathPhys2006, NestPRA2008}. On the other hand, frustration-free, noncommuting models include the Affleck-Kennedy-Lieb-Tasaki (AKLT) \cite{AffleckPRL1987}, Rokhsar-Kivelson models \cite{RokhsarPRL1988, CastelnovoAnnPhys2005} and parent Hamiltonians that are defined from injective projected entangled-pair states (PEPS) \cite{Perez-GarciaQIC2007, PerezGarciaQIC2008, SchuchAnnPhys2010, CiracRMC2019, MScCruz}. Sufficient conditions on when a Hamiltonian must be frustration-free have been studied in \cite{SattathPNAS2016}.
    \end{itemize}
    \item If $H$ is a sum of mutually commuting terms, its eigenstates correspond to eigenstates of each of the $H_i$. Note, however, that the order of the eigenenergies in $H$ needs not correspond to the order of the eigenenergies in $H_i$. For instance, changing $H_i$ to $-H_i$ reverses the order of the eigenstates, but leaves commutativity untouched. The simplest example of a commuting, non-frustration-free Hamiltonian is to consider $H = \sum_{<i,j> \in E} \sigma^{(i)}_z\otimes \sigma^{(j)}_z$, where $E$ are the edges of a triangle. In this case, tightening the constraints in \eqnref{eq:SdPCustom} helps in better capturing the frustration in the model, thus improving $\beta_C$, as a larger number of sites is considered. %
\end{itemize}

\section{Generalizations}
\label{sec:Generalization}
The framework that is here presented applies to the meta-problem of obtaining the best certificates given a computational budget by finding the most suitable convex relaxation. Although our case of study was centered around lower-bounding the ground state energy of local Hamiltonians, our methodology can be directly applied to many other tasks. The only requirement is to adapt the black box routine from \eqnref{eq:SdPCustom} to the new tasks, and appropriately map the constraint space to the new problem. Once it is done, provided that the presented optimization framework is entirely agnostic to the actual problem, its implementation to other tasks is straightforward.

Convex sets arise naturally in quantum information in many flavors. An efficient way to characterize them is through linear witnesses. Among those, witnesses that can be easily measured are clearly preferred. This property means, in practice, that they consist of an $O(\mathrm{poly}(n))$ number of terms. An important subclass of them is that in which these terms are local; \ie acting on $O(1)$ parties at most. In \appref{app:Generalization} we thoroughly discuss how to perform the SdP formalization of some relevant problems in quantum information. In \appref{app:EWs} we discuss an important class of entanglement witnesses, which are derived from local Hamiltonians, in \appref{app:NPA} we discuss how our approach can be used to optimize outer approximations to the set of quantum correlations, in \appref{app:SoSPoly} we consider the more general problem of finding better sum-of-squares representations of multivariate polynomials and in \appref{app:NLDepth} we discuss how our method can be applied in problems that are amenable to linear programming, such as finding outer approximations to projections of the set of correlations that satisfy the no-signalling principle.

\section{Conclusion and outlook}
\label{sec:concl}

In this work, we have introduced a novel approach to construct optimal relaxations to obtain certificates of quantum many-body properties, given a finite computational budget. Then, we have proposed a machine learning approach, based on deep reinforcement learning, to find such certificates. We have showcased its properties in the context of approximating the ground state energy of quantum local Hamiltonians. 

With the proposed framework, the RL agent is able to find the certificate that maximizes the objective function with the lowest complexity and whose cost lies within the computational budget. We have studied the validity of the method in the well-known Heisenberg XX model, showing that the agent is able to correctly characterize the ground state across the phase diagram. Indeed, we have shown how the certificates found by the agent change accordingly to the changes in the ground state.

Already for small systems, the agent is able to capture the complexity of the system of study and go beyond more trivial and simpler solutions, even when these are close in terms of the objective function. We have also shown that the agent is able to solve the opposite case, in which simpler proofs provide better bounds than more complex ones. Besides, we have shown that the qualitative solutions obtained in reduced systems can be used in larger ones, as these remain consistent for any size. Hence, the constraint optimization can be performed in a reduced version of the original problem in order to minimize the computational workload. 

Additionally, we have shown that the reinforcement learning approach handles large optimization spaces rather successfully, strongly outperforming other classical optimization algorithms. As final result, we have shown how to leverage transfer learning, positively impacting scalability. Moreover, we have characterized its behaviour, to find that it may be indicative of changes in the nature of the ground state of the system of study, some of which are due to phase transitions. The structure of the constraints which suffice for a good approximation correlates with the system's phase and the entanglement properties of the ground state. Unravelling their precise relation is a matter deserving future investigation.

Finally, we have provided an analysis of some particular cases within the context of ground energy estimation, as well as the tools to generalize the framework to other common tasks such as entanglement witnessing or outer approximations to the quantum set of correlations, to name a few. The presented framework can be readily extended to other tasks in quantum information that are based on finding good outer approximations of convex sets that are hard to describe.

As future work, it remains open the question of which properties of the Hamiltonian have lead to better bounds with cheaper solutions. Furthermore, transfer learning can be used to analyze common patterns between different Hamiltonians. Besides, the architecture of the reinforcement learning agent can be adapted to allow for the transfer learning between problems of different sizes. As an additional step, it would be interesting to study how introducing explicit information about the Hamiltonian may affect the optimization process.
For instance, whether a RL agent can help in designing better adiabatic schedules \cite{Schiffer2021} or whether better certificates can be built by combining RL following an adiabatic path.

\section{Code availability}

The code for the method proposed in this work is accessible in Ref.~\cite{requenazenodo2021} in form of a Python library, with tools to reproduce the results presented and use the method in various scenarios.

\section{Acknowledgements}
The authors acknowledge the contribution of Aina Guirao to the design of the figures. B.R., G.M.-G. and M.L. acknowledge support from ERC AdG NOQIA, Agencia Estatal de Investigaci\'on (“Severo Ochoa” Center of Excellence CEX2019-000910-S, Plan National FIDEUA PID2019-106901GB-I00/10.13039 / 501100011033, FPI), Fundaci\'o Privada Cellex, Fundaci\'o Mir-Puig, and from Generalitat de Catalunya (AGAUR Grant No. 2017 SGR 1341, CERCA program, QuantumCAT \_U16-011424, co-funded by ERDF Operational Program of Catalonia 2014-2020), MINECO-EU QUANTERA MAQS (funded by State Research Agency (AEI) PCI2019-111828-2 / 10.13039/501100011033), EU Horizon 2020 FET-OPEN OPTOLogic (Grant No 899794), and the National Science Centre, Poland-Symfonia Grant No. 2016/20/W/ST4/00314. G.M.-G. acknowledges funding from  Fundaci\'o Obra Social ``la Caixa'' (LCF-ICFO grant). J. T. thanks the Alexander von Humboldt foundation for support. This project has received funding from the Deutsche Forschungsgemeinschaft (DFG, German Research Foundation) – Project number 414325145 in the framework of the Austrian Science Fund (FWF): SFB F7104. This project has received funding from the European Union’s Horizon 2020 research and innovation programme under grant agreement No 899354.
This work was supported by the Dutch Research Council (NWO/OCW), as part of the Quantum Software Consortium programme (project number 024.003.037).
We thank A. Ac\'in, F. Alet, F. Baccari, M. Lubasch and N. Pancotti for enlightening discussions.

\bibliography{mylib}

\appendix
\section{SdP-based certificates}
\label{app:SdPBasics}
In this section we discuss the details on how a proof is obtained via a semidefinite program (SdP). To this end, let us recall the (primal) form of a SdP in canonical form:
\begin{equation}
    \begin{array}{llr}
    &\min_{X}& \langle C, X \rangle\\
    &\mathrm{s.t.}&\langle A_i, X\rangle = b_i\\
    &&X\succeq 0.
    \end{array}
    \label{eq:SdPPrimal}
\end{equation}
To each primal SdP one can associate a dual SdP, which is the following optimization problem.
\begin{equation}
    \begin{array}{llr}
    &\max_{\mathbf{y}}& \mathbf{y}^t \mathbf{b}\\
    &&C - \sum_i y_i A_i \succeq 0.
    \end{array}
    \label{eq:SdPDual}
\end{equation}
Although the primal SdP \eqnref{eq:SdPPrimal} and the dual SdP \eqnref{eq:SdPDual} are different optimization problems any two $X$ and $\mathbf{y}$ that satisfy their respective problem constraints obey the weak duality relation
\begin{equation}
    \langle C, X \rangle  \geq \mathbf{y}^t \mathbf{b}.
    \label{eq:weakduality}
\end{equation}
This means that a feasible solution $X$ of \eqnref{eq:SdPPrimal} upper bounds the value of the objective function of \eqnref{eq:SdPDual} (in particular, its maximum value). Conversely, any feasible solution $\mathbf{y}$ of \eqnref{eq:SdPDual} lower bounds the value of the objective function of \eqnref{eq:SdPPrimal} (in particular, its minimum value). Therefore, one can construct a mathematical proof for a bound $E_0 \geq \beta$ by transforming the SdP into primal canonical form \eqnref{eq:SdPPrimal}, constructing its dual problem and finding a dual feasible solution of the latter \eqnref{eq:SdPDual}. In practice, this transformation is taken care of automatically with SdP parsers such as cvx \cite{CVX1, CVX2} or yalmip \cite{YALMIP} and an SdP solver (e.g. \cite{SeDuMi, SDPT3}) numerically finds the values of the optimal $X$ and $\mathbf{y}$ of both problems. Furthermore, \eqnref{eq:weakduality} typically becomes tight at optimality of both \eqnref{eq:SdPPrimal} and \eqnref{eq:SdPDual} under reasonable regularity conditions, such as strict feasibility~\cite{ParriloBook2013}.

\section{Pathological cases}
\label{app:Pathological}
It might appear that a more natural way to define \eqnref{eq:SdPCustom} would be as follows:
\begin{equation}
    \displaystyle
    \label{eq:SdPCustomNaive}
    \begin{array}{llrr}
    \beta_{C}:=&\min_{\Xi_C}& \sum_i \langle H_i \rangle&\\
    &\mathrm{s.t.}&\rho_{S} \succeq 0 & \forall S \in C\\
    &&\mathrm{Tr}[\rho_S] = 1&\\
    &&\mathrm{Tr}_{R^c}[\rho_S] = \mathrm{Tr}_{R^c}[\rho_{S'}],&\mbox{where } R = S \cap S',\quad \forall S, S' \in C.
    \end{array}
\end{equation}

\begin{figure}
    \centering
    \includegraphics[width=0.3\textwidth]{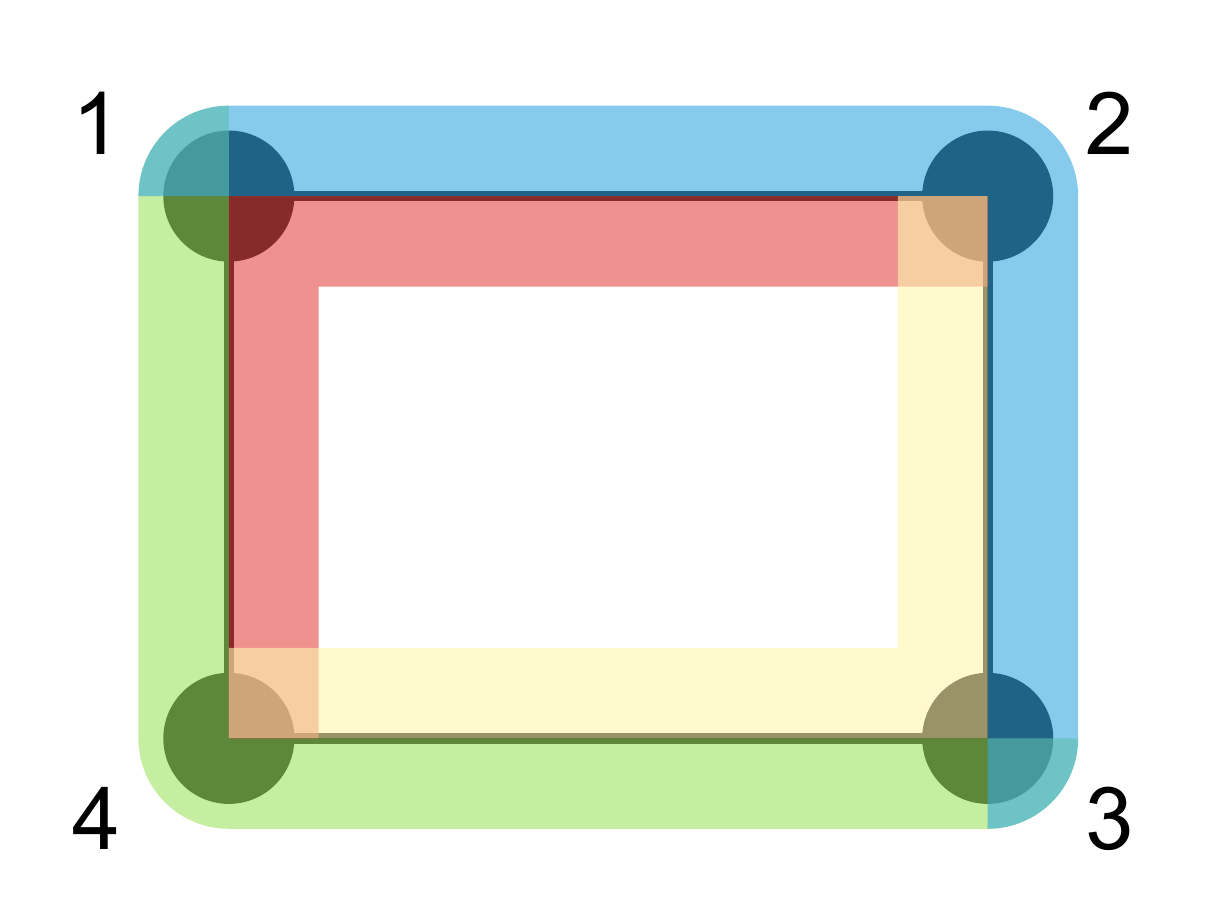}
    \caption{A pathological case for \eqnref{eq:SdPCustomNaive}.}
    \label{fig:pathological}
\end{figure}

In other words, \eqnref{eq:SdPCustomNaive} enforces that reduced states are equal in every pairwise intersection of constraints. Both \eqnref{eq:SdPCustom} and \eqnref{eq:SdPCustomNaive} yield valid certificates, but \eqnref{eq:SdPCustomNaive} might not implement all the compatibility conditions that one would naively expect, in some pathological cases. Here we discuss an example (see \figref{fig:pathological}). Consider a system of $4$ qubits in a $1-$D geometry on a ring \ie, with periodic boundary conditions. Consider furthermore that our set of constraints is $C=\{\{0,1,2\},\{1,2,3\},\{2,3,0\},\{3,0,1\}\}$. \eqnref{eq:SdPCustomNaive} would require that the SdP takes into account the variables $\rho_{0,1}$, $\rho_{0,2}$, $\rho_{0,3}$, $\rho_{1,2}$, $\rho_{1,3}$ and $\rho_{2,3}$. These come from the $6$ ways to choose $2$ elements from a $4$-element set, like $C$. Note, however, that the SdP does not enforce conditions that one would naturally expect, such as
$$\rho_3 \equiv \mathrm{Tr}_{2}[\rho_{23}] = \mathrm{Tr}_{0}[\rho_{03}] = \mathrm{Tr}_{1}[\rho_{13}].$$
In other words, the two-body reduced density matrices stemming from \eqnref{eq:SdPCustomNaive} do not need, \textit{a priori}, to have compatible supports in their intersections. This caveat is resolved in the formulation of \eqnref{eq:SdPCustom}. In practice, however, pathological cases such as the one depicted in \figref{fig:pathological} are quite rare. For instance, the same scheme with a three-body constraint centered at each site, but for a number of parties larger than $4$ would automatically generate all the single-body terms.

\section{Optimal constraints across system sizes}
\label{app:SolsLargerN}

In \secref{sec:opt_results} we presented the result of applying the proposed method to to the XX Heisenberg model~\eqnref{eq:XX} with a computational budget that allowed the agent to allocate up to half of the 3-body constraints. Here, we explore in further detail how the different sets depicted in~\figref{fig:toy_examples} vary in energy throughout the phase space. Moreover, we show that even for bigger system sizes the same qualitative solutions remain optimal across the different sizes. In particular, we will study systems of sizes $n=6, 12, 24, 36$.

Let us first address what we mean by qualitative solutions and how these are generalized to different sizes. The optimal sets of constraints shown in~\figref{fig:toy_examples} can be seen as patterns of reduced density matrices (RDMs) that span the system and can therefore be reproduced at any size. This way, sets of constraints made out of the same RDM pattern may be seen as the same qualitative solution. Let us describe these patterns and provide some examples:
\begin{description}
\item[(a)] Span the system with evenly distributed 3-body RDMs.
\begin{itemize}
    \item $n=6$: $C=\{\{0, 1, 2\}, \{2, 3, 4\}, \{4, 5, 0\}\}$, $C=\{\{1, 2, 3\}, \{3, 4, 5\}, \{5, 0, 1\}\}$,
    \item $n=12$: $C=\{\{0, 1, 2\}, \{2, 3, 4\}, \{4, 5, 6\}, \{6, 7, 8\}, \{8, 9, 10\}, \{10, 11, 0\}\}$.
\end{itemize}
\item[(b)] Span the system with 3-body RDMs, including an additional 2-body RDM that shifts one of the 3-body ones. This is the least straightforward pattern to generalize, provided that it can either be interpreted as having only one extra 2-body RDM, or including some additional 2-body RDMs every few sites. We have found that, for the considered system sizes, the optimal set of constraints is found by including these 2-body RDMs every $6$ sites, \ie spanning the system by repetition of the $6$-body pattern. 
\begin{itemize}
    \item $n=6$: $C=\{\{0, 1, 2\}, \{2, 3, 4\}, \{3, 4, 5\}, \{5, 0\}\}$, $C=\{\{0, 1, 2\}, \{2, 3\}, \{3, 4, 5\}, \{4, 5, 0\}\}$,
    \item $n=12$: $C=\{\{0, 1, 2\}, \{2, 3, 4\}, \{3, 4, 5\}, \{5, 6\}, \{6, 7, 8\}, \{8, 9, 10\}, \{9, 10, 11\}, \{11, 0\}\}$.
\end{itemize}
\item[(c)] Span the system alternating 3-body and 2-body RDMs. 
\begin{itemize}
    \item $n=6$: $C=\{\{0, 1, 2\}, \{2, 3\}, \{3, 4, 5\}, \{5, 0\}\}$, $C=\{\{0, 1\}, \{1, 2, 3\}, \{3, 4\}, \{4, 5, 0\}\}$,
    \item $n=12$: $C=\{\{0, 1, 2\}, \{2, 3\}, \{3, 4, 5\}, \{5, 6\}, \{6, 7, 8\}, \{8, 9\}, \{9, 10, 11\}, \{11, 0\}\}$.
\end{itemize}
\item[(d)] Span the system with 2-body RDMs. 
\begin{itemize}
    \item $n=6$: $C=\{\{0, 1\}, \{1, 2\}, \{2, 3\}, \{3, 4\}, \{4, 5\}, \{5, 0\}\}$, $C=\{\{1, 2\}, \{2, 3\}, \{3, 4\}, \{4, 5\}, \{5, 0\}\}$,
    \item $n=12$: $C=\{\{0, 1\}, \{1, 2\}, \{2, 3\}, \{3, 4\}, \{4, 5\}, \{5, 6\}, \{6, 7\}, \{7, 8\}, \{8, 9\}, \{9, 10\}, \{10, 11\}, \{11, 0\}\}$.
\end{itemize}
\end{description}

\begin{figure}
    \centering
    \begin{subfigure}{.5\textwidth}
      \centering
      \includegraphics[width=\textwidth]{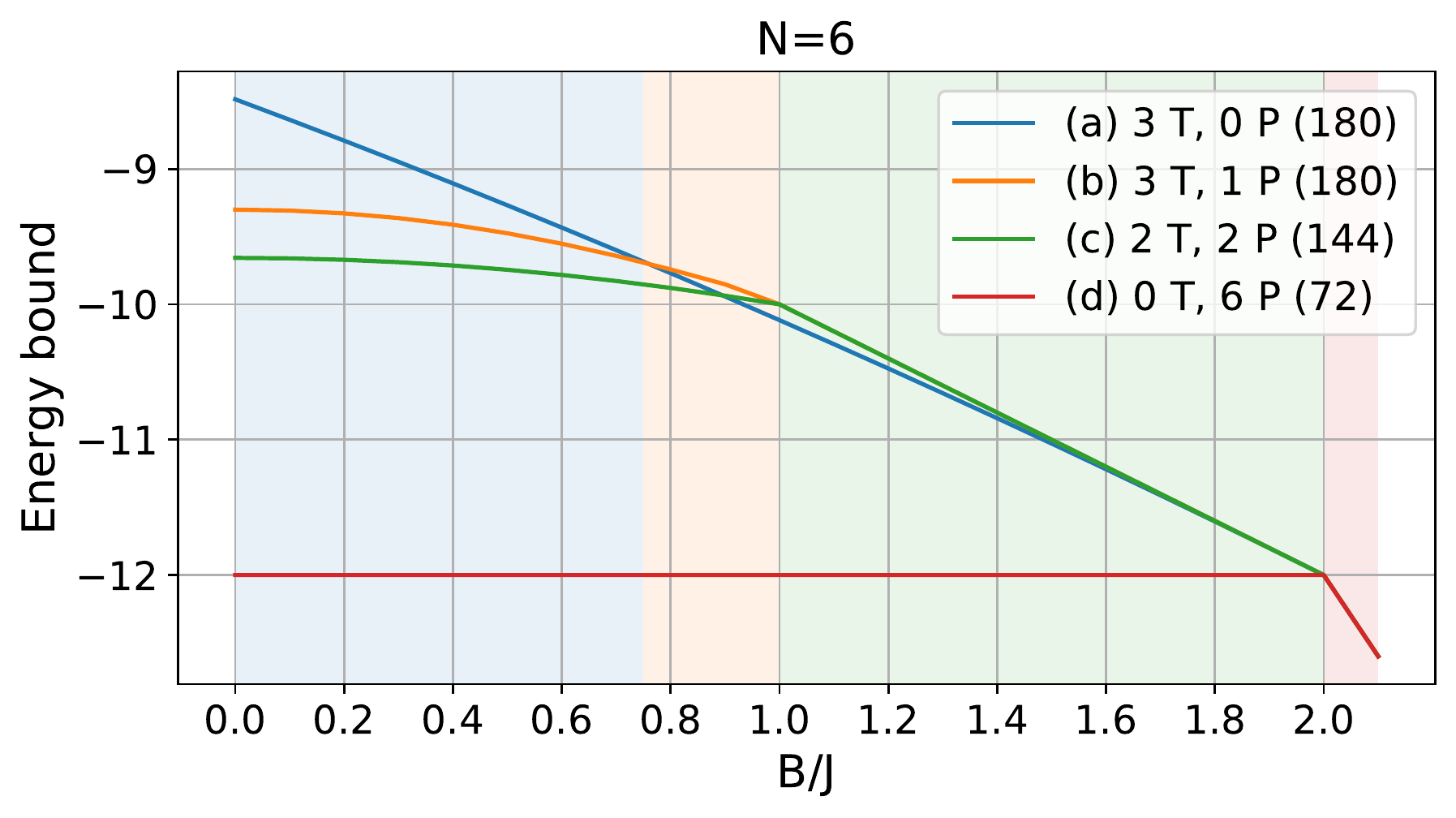}
      \caption{Energy bounds for $n=6$.}
    \end{subfigure}%
    \begin{subfigure}{.5\textwidth}
      \centering
      \includegraphics[width=\textwidth]{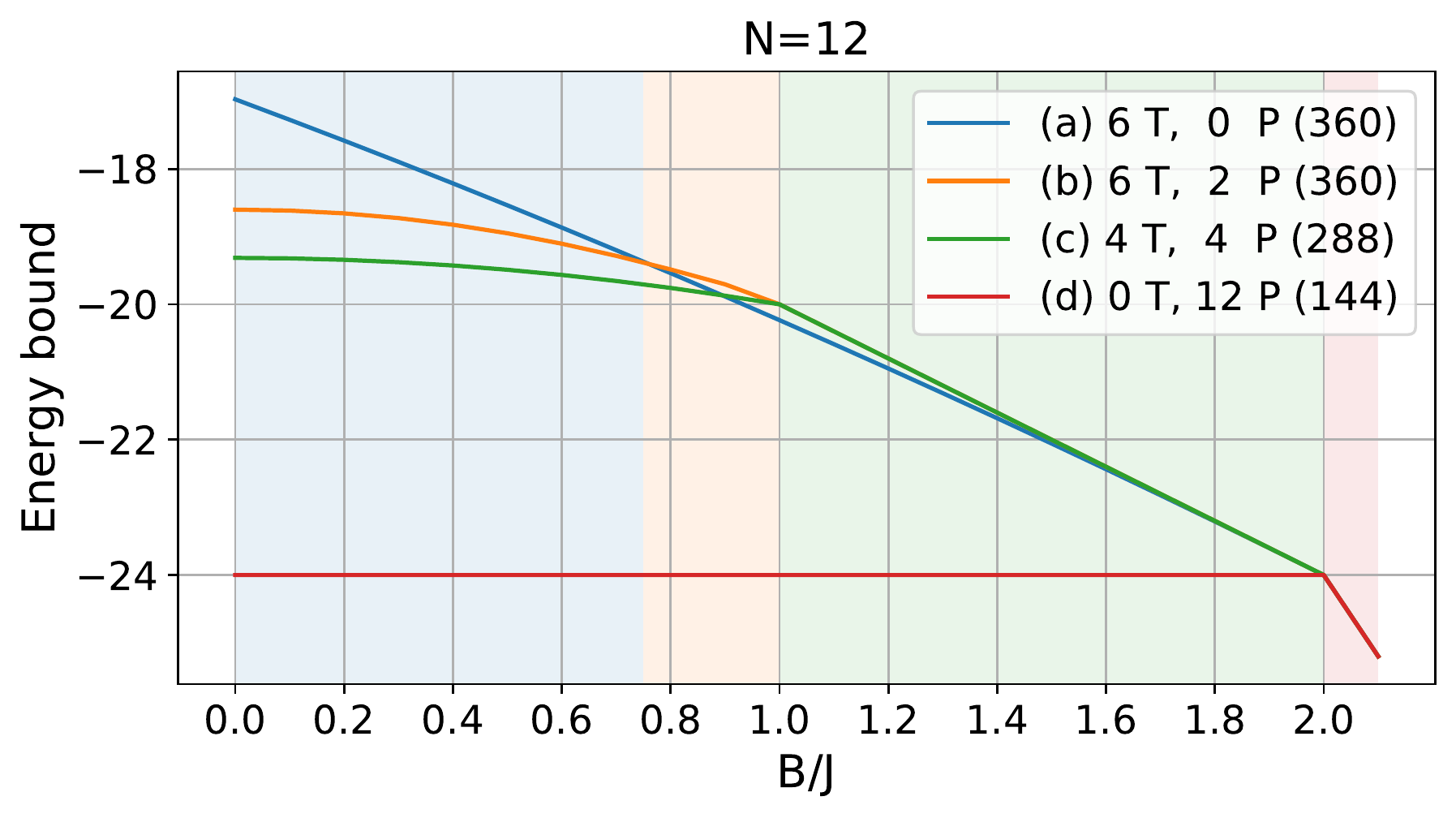}
      \caption{Energy bounds for $n=12$.}
    \end{subfigure}
    \begin{subfigure}{.5\textwidth}
      \centering
      \includegraphics[width=\textwidth]{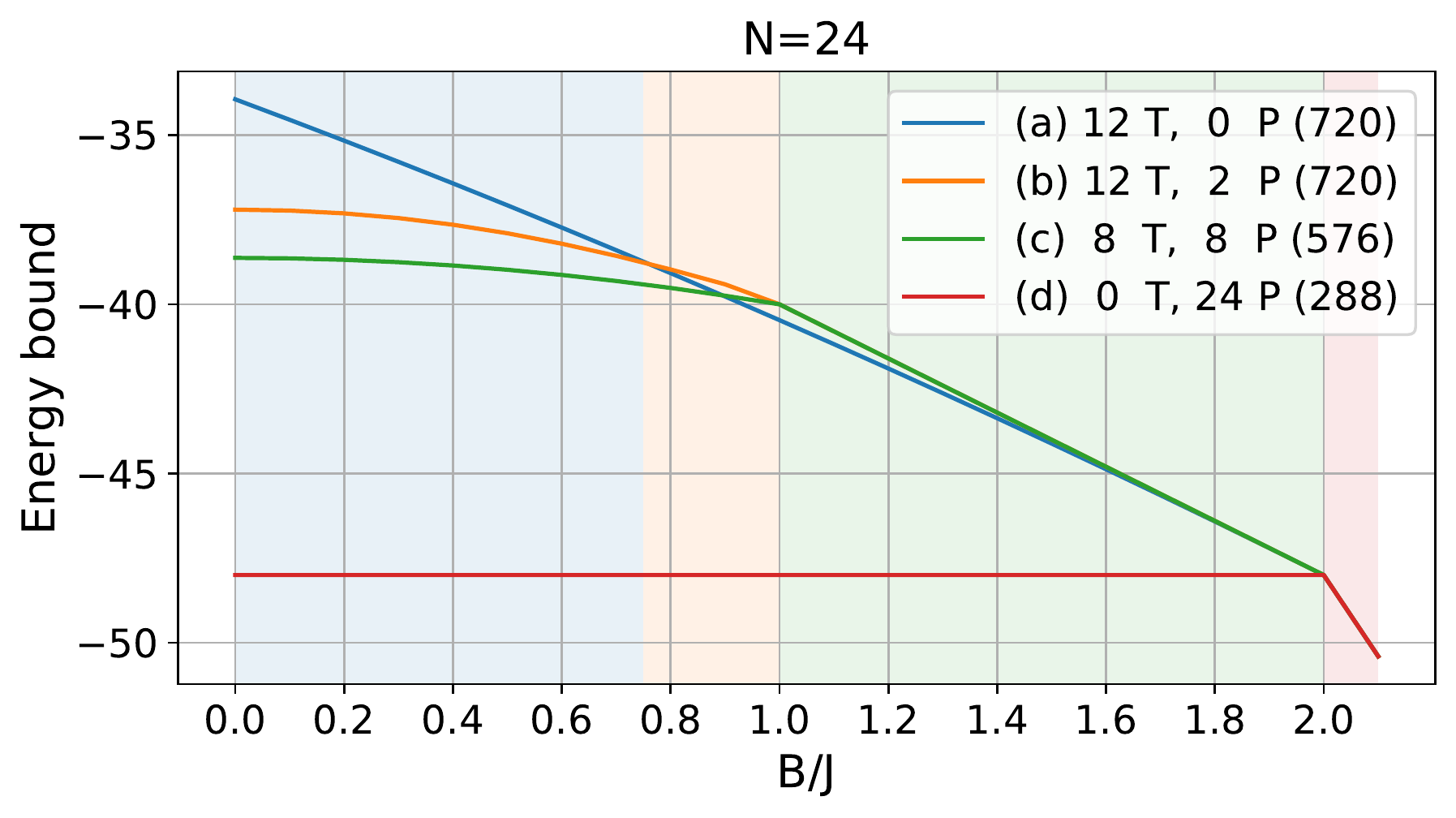}
      \caption{Energy bounds for $n=24$.}
    \end{subfigure}%
    \begin{subfigure}{.5\textwidth}
      \centering
      \includegraphics[width=\textwidth]{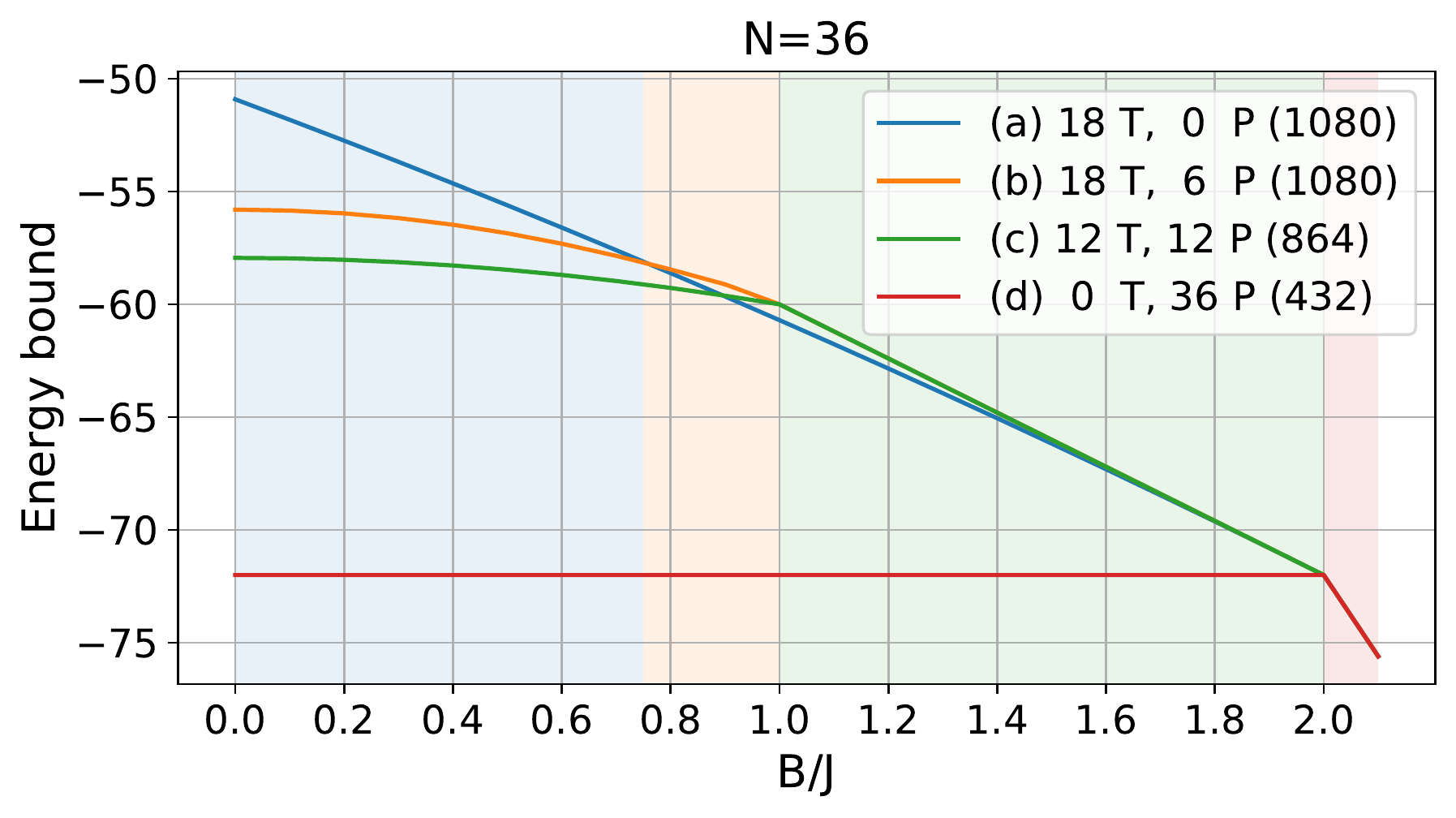}
      \caption{Energy bounds for $n=36$.}
    \end{subfigure}%
    \caption{Energy bounds obtained for system sizes of $n=6, 12, 24, 36$ with the sets of constraints that are optimal in some region of the phase space. The labels (a), (b), (c), (d) refer to the qualitative solutions of~\figref{fig:toy_examples}, followed by the number of triplets (T) and pairs (P) constituting the RDMs and, in brackets, the cost associated to solving the associated SdP in terms of the number of free variables in the SdP. The shaded background indicates the color of the set of constraints that is optimal within the range.}
    \label{fig:optimal_sizes}
\end{figure}

In~\figref{fig:optimal_sizes} we show the energy bounds obtained by all the sets of constraints that, at some point along the phase diagram, are optimal. Indeed, we find that the optimal sets of constraints at different system sizes represent the same qualitative solutions and the regimes under which these are optimal are all the same. This suggests that a reduced version of the original problem can be used in order to find the optimal set of constraints, significantly reducing the computational cost of the optimization.

Additionally, we see that the qualitative solution (c), in some regions of the phase space, provides the same energy bound as (b) and, even more, yields a better bound than (a), while being a much simpler certificate that involves $20\%$ less SdP variables than (a) and (b). Overall, the relative behaviour of each set of constraints is the same across system sizes and they all converge to the same value at $B/J=2$, where the phase transition happens.  

\section{Optimization details}
\label{app:Hyperparams}
In this section, we provide a brief description of the RL method, with details about the specific parameters involved in each one to obtain the results shown in \figref{fig:benchmark}. These are the results of finding the optimal set of constraints for various system sizes and two different computational budgets. Moreover, we also present the parameters needed to reproduce the results of the breadth first search (BFS)~\cite{CormenBook2009} and Monte Carlo~\cite{KirkpatrickScience1983} approaches used in the same figure.

\subsection{Reinforcement learning parameters}

The RL optimization has several hyper-parameters that dictate both the deep Q-network architecture~\cite{MnihNature2015} and the learning procedure of the RL agent. Given that the size of the state vectors depends on the size of the actual problem (see \secref{sec:constraintopt}), the RL agent must be adapted to each system size. This way, we define many of the parameters as a function of the system size $n$, the state vector size $s$ and the number of possible actions $a=s+1$, as the agent is allowed to remain in the same state. 

The agent architecture has three fully connected hidden layers with rectified linear unit (ReLU) activation function. The input layer has size $s$, the first layer has size $3s$, the second layer has size $2a$, the third layer has size $2a$ and the output layer has size $a$. This network is copied as a target network for double Q-learning~\cite{vanHasseltAAAI2016}.

The learning procedure is structured in learning episodes in which the agent performs a trajectory through the state space, always starting from the same initial state, as described at the beginning of \secref{sec:constraintopt}. Throughout the learning episodes, the agent gathers experience in the form of \textit{State-Action-New State-Reward} tuples that are stored in a memory. At the end of each episode, the agent replays a batch of steps from the memory to learn. 

We set an episode length of the order of the system size $n$, modified according to the computational budget. For a low budget, such as in \figref{fig:bench_half} with half of the 3-body constraints available, the episode length can be slightly lower than $n$, \eg $0.7n$. In contrast, for a high budget, as in \figref{fig:bench_all} with all the 3-body constraints available, the episode length needs to be slightly higher than $n$, \eg $1.2n$. This way, we guarantee that the agent has enough time to allocate all the possible constraints with some margin for errors. We set the batch size for the experience replay to $20$ episodes and the agent starts learning once it has visited as many states as a quarter of the batch size. We use a learning rate of $5\times 10^{-3}$ and we update the target network every $5$ episodes.

Finally, in order to enforce the agent to explore, we vary the value of $\epsilon$ in the $\epsilon-$greedy policy throughout the training process. We start with $\epsilon_0=0.9$ and we make it decay exponentially after every training episode, labelled by $e$, such that $\epsilon_{e}=\max\left\{0.1, \delta^{e}\epsilon_0\right\}$ with $\delta\in(0, 1)$. In the benchmarking from \figref{fig:benchmark}, we have taken $\delta=0.5$ for systems $n\leq7$ and $\delta=0.95$ for systems $n>7$. In small systems, the constraint space is reduced and the exploration is not needed, while a proper exploration is critical in larger problems.

\subsection{Benchmarking optimization methods}

Here we describe the methods used to benchmark our results: bread first search (BFS) and Monte Carlo (MC) optimization.

BFS does not have any hyper-parameter. Starting from the initial state, it builds a queue of states to visit by recursively expanding each state. Expanding a state consists on appending, at the end of the queue, all the possible states that can be reached from it through valid actions. We have taken randomized orders in the state expansion and we do not consider states that have already been visited or that are already in the queue. 

The MC optimization has only one hyper-parameter, that is, the effective temperature $T$. The algorithm consists on proposing random valid actions to go from one state to another. Then, the movement is accepted or rejected depending on the reward associated to the old and new states, $R_{\mathrm{old}}, R_{\mathrm{new}}$, with acceptance probability $p(R_{\mathrm{new}},R_{\mathrm{old}})=\min\left\{1, e^{(R_{\mathrm{new}}-R_{\mathrm{old}})/T}\right\}$. We tune the effective temperature to obtain a $50\%$ acceptance ratio in a long, well converged, optimization. The results from \figref{fig:bench_half} are obtained with $T=0.084$ and the results from \figref{fig:bench_all} with $T=0.097$. 

\section{Generalizations}
\label{app:Generalization}
In \secref{sec:Generalization}, we briefly mention the generalization of the presented method to other common tasks in the field of quantum information processing, beyond the running example of lower bounding the ground state energy of local Hamiltonians. We present, here, a non-exhaustive set of examples explicitly showing how to implement the black box routine from \eqnref{eq:SdPCustom} to such tasks. This allows the straightforward implementation of the RL framework, provided that is entirely agnostic to the actual problem.

\subsection{Entanglement witnesses from local Hamiltonians}
\label{app:EWs}
The most straightforward adaptation of our case study relates to entanglement detection \cite{GuehnePhysRep2009}. In particular, to finding relaxations to the separability bound of entanglement witnesses constructed from local Hamiltonians \cite{TothApplPhysB2005, GuehneNJP2005}. In this case, we need ask that the global quantum state $\rho$ is fully separable. Note that the global optimization task is to minimize $\langle H \rangle$ over the set of separable states, which form a convex set. The RDMs of a fully separable $\rho$ are also separable, but the separability condition is hard to impose in a SdP. Actually, deciding membership in the set of separable quantum states has been shown to be NP-hard \cite{Gurvits2003}, even in notably simpler instances \cite{YuPRA2016, TuraQuantum2018, Marconi2020}. However, the set of separable states is contained in the set of states that fulfill the PPT condition, which are easy to characterize via a SdP. Hence, \eqnref{eq:SdPCustom} can be straightforwardly modified to yield a lower bound on the separable bound of a local Hamiltonian $H$, when $H$ is seen as an entanglement witness:

\begin{equation}
    \displaystyle
    \label{eq:SdPEW}
    \begin{array}{llrr}
    \beta_{C}^{\mathrm{full-sep}}:=&\min_{\Xi_C}& \sum_i \langle H_i \rangle&\\
    &\mathrm{s.t.}&\rho_{S} \succeq 0 & \forall S \in C\\
    &&\rho_S^{\Gamma_A} \succeq 0&\forall A \subseteq S\\
    &&\mathrm{Tr}[\rho_S] = 1&\\
    &&\mathrm{Tr}_{R^c}[\rho_S] = \mathrm{Tr}_{R^c}[\rho_{S'}]&\forall R \subseteq S \cap S',\quad S, S' \in C,
    \end{array}
\end{equation}
where the superscript $\Gamma_A$ indicates that the partial transposition $({\mathbbm{1}}_{A^c}\otimes T_{A})$ has been applied to the elements of $S = A \cup A^c$. Note that this is a linear operation since it simply permutes  elements of $\rho_S$. Hence, any quantum state satisfying $\mathrm{Tr}[\rho H] < \beta_C^{\mathrm{full-sep}}$ contains some entanglement.

The optimization in \eqnref{eq:SdPEW} can be tightened in several directions. First, one can consider symmetric extensions \cite{DohertyPRA2004} in order to improve the approximation of the PPT set to the separable set, at a cost of increasing the computational demands of \eqnref{eq:SdPEW}. In some cases, one can furthermore demand that the bound detects a higher degree of entanglement, yielding a $k$-producibility bound. In \cite{AloyPRL2019, TuraPRA2019} device-independent witnesses of entanglement depth have been proposed, and their lower bounds are found via a SdP that encodes a relaxation of the quantum marginal problem. In this case, the relaxation can be tightened by imposing compatibility with larger quantum states, as long as these remain within a computational budget.

\subsection{Outer approximations to the set of quantum correlations}
\label{app:NPA}
The set of correlations that are produced by quantum mechanics is also a convex set \cite{BrunnerRMP2014}. A whole program aiming at its characterization has obtained several operationally-motivated characterizations of it \cite{PR94, BrassardPRL2006, LindenPRL2007, NavascuesPRSA2010, PawlowskiNature2009, FritzNatComms2013, GallegoPRL2011, NavascuesNatComms2015}. Systematic methods also yield relaxations, which can be made arbitrarily accurate at a higher computational cost \cite{NavascuesPRL2007, NavascuesNJP2008, PironioSIAM2010}.
In this case, we note we also have a poset structure that can be exploited to build a similar constraint space.

Let us recall that the so-called Navascu\'es-Pironio-Ac\'in (NPA) hierarchy \cite{NavascuesPRL2007} chooses a set of operators $S = \{\mathbbm{1}, A_0, B_0, \ldots\}$ from which it builds a moment matrix $\Gamma = S^\dagger S$. Non-trivial relationships among the entries of $\Gamma$ are imposed by the algebra generated by the elements of $S$: commutation relations or identities such as $A_i^\dagger A_i = \mathbbm{1}$ impose linear constraints among the entries of $\Gamma$. Then, given a Bell inequality that can be formally represented as $I = \mathrm{Tr}[C \Gamma]$, one can find a lower bound to its value over the quantum set by solving
\begin{equation}
    \displaystyle
    \label{eq:NPA}
    \begin{array}{llrr}
    \beta_{Q}^{S}:=&\min_{\Gamma}& \mathrm{Tr}[C \Gamma]&\\
    &\mathrm{s.t.}&\Gamma \succeq 0 &\\
    &&\Gamma_{00} = 1&\\
    &&\mathrm{Tr}[{C_i \Gamma}] = 0&,
    \end{array}
\end{equation}
where $C$ and $C_i$ are real matrices, thus obtaining a quantum Bell inequality of the form $I \geq \beta^S_Q$. Note that $C$ picks the coefficients of $I$ and the $C_i$ enforce the conditions arising from the operator algebra. For instance, if the Bell scenario is such that the outcomes of the measurements are $\pm 1$ then $A_k^2 - \mathbbm{1} = 0$. Then $C_i$ picks the entries $A_k^2$ and $\mathbbm{1}$ in $\Gamma$ with the appropriate coefficients, imposing the equality constraint. Similarly, commutation relations such as $[A_k, B_l] = 0$ are enforced in the same way.

In this case, the poset structure lies in the definition of the set of operators $S$. The partial order relation $\preccurlyeq$ corresponds to the inclusion order relation $\subseteq$ between two different sets of operators and the agent can perform actions in a similar way, by adding and removing operators.

In analogy to \secref{sec:Particularcases}, some witnesses for the quantum set admit a proof for a ver low operator degree in $S$. The paradigmatic example is the CHSH Bell inequality \cite{CHSH}, which can be shown to be bounded by $2\sqrt{2}$:
\begin{equation}
    2\sqrt{2}\mathbbm{1} - (A_0B_0 + A_0 B_1 + A_1 B_0 -A_1B_1) = \frac{1}{\sqrt{2}}\sum_{i=0}^1\left(A_i - \frac{B_0 +(-1)^i B_1}{\sqrt{2}}\right)^\dagger\left(A_i - \frac{B_0 +(-1)^i B_1}{\sqrt{2}}\right) \succeq 0.
    \label{eq:CHSHsos}
\end{equation}

On the other hand, inequalities such as the so-called $I_{3322}$ inequality \cite{FroissartIlNuovoCimento1981} do not seem to admit a tight proof for their quantum bound, even in the case that $S$ contains operators up to degree $5$ \cite{PalPRA2010, PhDRosset}. Having simple certificates such as those of the form of \eqref{eq:CHSHsos} turns out to be extremely convenient for proofs in device-independent quantum information processing protocols, such as self-testing: When $\beta^S_Q$ is tight, it means that the quantum state and measurements yield exactly zero expectation value on all the sos terms (cf. \eqref{eq:CHSHsos}). These equations then impose conditions that allow to characterize the states and/or measurements performed to some extent, solely from their statistics \cite{Supic2019, AcinPRL2012, YangPRAR2013, BampsPRA2015, ColadangeloNatComms2017, SupicNJP2018, KaniewskiQuantum2019, BaccariPRL2020}.

\subsection{Improving sum-of-squares representations of non-negative polynomials}
\label{app:SoSPoly}

Semi-definite programming optimization is essentially equivalent to finding sum-of-squares decompositions \cite{ParriloBook2013}. The latter arise naturally when trying to answer the following question: given a real polynomial in $d$ variables, does it take non-negative for all points in $\mathbbm{R}^d$? This is precisely Hilbert's $17$th problem \cite{HilbertProblems1902}. On the one hand, it is a trivial observation that every polynomial that admits a sum-of-squares representation is non-negative by construction, and the latter can be efficiently found via a SdP. Unfortunately, not every non-negative polynomial admits a sum-of-squares decomposition in terms of polynomials \cite{Motzkin1967}. In fact, although Hilbert's problem was solved by Artin in 1927 \cite{Artin1927}, who showed that every non-negative polynomial admits a sum-of-squares representation in terms of rational functions, the problem remains NP-hard. By controlling the degree of the denominator in the rational function sos, one also obtains a hierarchy.

Interestingly, non-negative polynomials also appear naturally in physics and optimization. For instance, imagine one wants to find the minimal energy of a classical local Hamiltonian. This task appears naturally in the verification of quantum optimizers \cite{BaccariPRR2020} or in the context of finding the classical bound of a Bell inequality with few-body correlators \cite{TuraPRX2017}. In these cases, there are some geometric properties imprinted in the cost function which one would like that they persist in the sos decomposition. However, sos decompositions are not unique in general. When the underlying graph that connects variables that interact directly is chordal, the sparsity in the objective function percolates to a sparse sos decomposition \cite{VandenbergheFTO2015, CifuentesSIAM2016, CifuentesSIAM2017, ZhengMatProg2019}. However, in the case that the underlying graph has a complicated chordal extension, it is significantly harder to obtain good sos decompositions, since there is no systematic method in this case, making the situation amenable to a RL agent. It would be interesting to see to which extent a RL agent recovers a perfect elimination ordering stemming from a chordal graph and whether it can find effective strategies when the graph is approximately chordal.

\subsection{Optimization of nonlocality depth witnesses from few-body Bell inequalities}
\label{app:NLDepth}

Here we consider the following multipartite Bell scenario, where $n$ parties labelled from $[n]$ are space-like separated and each of them can perform $m$ measurements each yielding $d$ possible outcomes. At the end of the experiment, parties have collected enough statistics to estimate the conditional probability distribution $p(\mathbf{a}|\mathbf{x})$, where $\mathbf{x}$ is an $n$-dimensional vector denoting a collective choice of measurements and $\mathbf{a}$ is also an $n$-dimensional vector labelling the corresponding outcomes. Studying Bell nonlocality in such a multipartite scenario easily turns into a highly complex task, even from the point of view of designing or finding relevant Bell inequalities \cite{WernerPRA2001, ZukowskiBruknerPRL2001, SliwaPLA2003, BancalJPA2010, SciencePaper}. Furthermore, in the multipartite scenario, analogously to entanglement \cite{AloyPRL2019, TuraPRA2019, GuehnePhysRep2009, LueckePRL2014, GuehneNJP2005}, Bell nonlocality can come in many flavors, from fully local models to bilocal models that are only falsified by genuinely nonlocal correlations. In addition, the multipartite Bell scenario poses the extra challenge of a consistent time ordering in defining a partially local model, otherwise it could be self-contradicting \cite{GallegoPRL2012, BancalPRA2013}. This caveat can be avoided by defining a so-called $k$-local model, which is a mixture of models of the form
\begin{equation}
    p(\mathbf{a}|\mathbf{x}) = \sum_{\lambda}p(\lambda) \prod_{i=1}^L p(\mathbf{a}_{S_i}|\mathbf{x}_{S_i}, \lambda),
    \label{eq:klocalmodel}
\end{equation}
where $\{S_i\}_{i=1}^L$ form a partition of $[n]$ with $|S_i|\leq k$, the so-called response functions $p({\mathbf{a}_{S_i}}|\mathbf{x}_{S_i},\lambda)$ satisfy the no-signalling principle and $\mathbf{a}_S$, $\mathbf{x}_S$ indicate that we select from $\mathbf{a}$ or $\mathbf{x}$, respectively, only those components whose index belongs to $S\subseteq[n]$. By mixing models of the form \eqnref{eq:klocalmodel}, one constructs $k$-local models, in this case, under no-signalling constraints.

In \cite{BaccariPRA2019, MScLopezPastor} a way to optimize Bell inequalities for $k$-nonlocality depth was proposed for large system sizes, leveraging on two factors that simplify the problem: designing Bell inequalities that are (i) permutationally invariant and (ii) composed of two-body correlators only. As can be inferred from \eqnref{eq:klocalmodel}, in order to construct these inequalities or to find their $k$-local bound given one, one needs to know a characterization, in terms of extremal points, of the projected no-signalling polytope for $|S_i|$ parties in the relevant Bell scenario. This way, one can construct the $k$-local polytope \cite{SliwaPLA2003, AnnPhys, BaccariPRA2019}.
Unfortunately, the polytope of nonsignalling correlations admits an easy description only in terms of inequalities. In terms of vertices, the so-called PR-boxes \cite{PR94}, it has been shown that finding all PR-boxes equivalent to finding all Bell inequalities \cite{FritzJMP2012}. Therefore, it seems that such a daunting task \cite{BabaiCompCompl1991, ChazelleDCG1993} could benefit from a relaxation approach, which we here describe: By allowing for a simpler characterization of the projected no-signalling polytope, one can hope to construct $k$-local models with less extremal points to be considered.

Let us recall that the no-signalling principle states that the probability distributions $p(\mathbf{a}|\mathbf{x})$, apart from satisfying the relations $\sum_\mathbf{a} p(\mathbf{a}|\mathbf{x}) = 1$ for all $\mathbf{x}$ and $p(\mathbf{a}|\mathbf{x})\geq 0$ for every $\mathbf{a}, \mathbf{x}$, as does any mathematically sound probability distribution, they also satisfy the so-called no-signalling principle, which reads
\begin{equation}
    p(\mathbf{a}_S|\mathbf{x}_S \cup \mathbf{x}_{S^c}) = p(\mathbf{a}_S|\mathbf{x}_S\cup \mathbf{x'}_{S^c}), \forall \mathbf{a}_S, \mathbf{x}_S, \mathbf{x}_{S^c}, \mathbf{x'}_{S^c}, S \subseteq[n]
\end{equation}
where we have split $\mathbf{x}$ into those components labelled in $S$, $\mathbf{x}_S$ and those in its complementary set, and $p({\mathbf{a}_S|\mathbf{x}})$ is defined as the marginal probability distribution
\begin{equation}
    p(\mathbf{a}_S|\mathbf{x}) = \sum_{a_i: i \in S^c} p(\mathbf{a}|\mathbf{x}).
\end{equation}
Note that, operationally, the NS principle imposes that the marginal probability distribution that a subset $S$ of parties observe does not depend on the inputs $\mathbf{x}_{S^c}$ received by the rest of the parties during the experiment. Hence, the rest of the parties cannot signal information to the parties in $S$ by choosing a particular sets of inputs. Furthermore, the no-signalling principle tells us that the quantities $p(\mathbf{a}_S|\mathbf{x}_S)$ are well defined.

We can now build the relaxation as follows: The projected no-signalling polytope, in terms of $2$-body correlations, is given in terms of the marginals $p(\mathbf{a}_S|\mathbf{x}_S)$, where $|S| = 2$ (we can take particular linear combinations of them to build symmetric correlators). Each of the $p(\mathbf{a}_S|\mathbf{x}_S)$ stemmed from a common $p(\mathbf{a}|\mathbf{x})$, but at a first relaxation level, this assumption can be dropped. The relaxation hierarchy is then built by imposing compatibility at larger and larger levels: for instance, given $S, W \subseteq [n]$, $|S|=|W|=2$, $|S\cap W| = 1$, we can impose that there exists a no-signalling  three-partite $p(\mathbf{a}_{S\cup W}|\mathbf{x}_{S\cup W})$ with appropriate marginals. From the set inclusion relation one recovers the same poset structure in the constraint space. Since now the problem is linear, one can build outer approximations to the projected no-signalling polytope by means of a linear programming black box or, equivalently, a diagonal SdP.

\end{document}